\documentclass{jfm}

\usepackage{graphicx}
\usepackage[table]{xcolor}
\usepackage{newtxtext}
\usepackage{newtxmath}
\usepackage{natbib}
\usepackage{hyperref}
\usepackage{xcolor}
\usepackage{caption}
\hypersetup{
    colorlinks = true,
    urlcolor   = blue,
    citecolor  = black,
}

\newcommand{\RomanNumeralCaps}[1]
\linenumbers

\title{Understanding the effect of wall elasticity in turbulent channel flows}

\author{Morie Koseki, M. S. Aswathy\footnote{Current address: Department of Aerospace Engineering, Indian Institute of Space Science and Technology, Thiruvananthapuram 695547, Kerala, India} \and Marco Edoardo Rosti \corresp{\email{marco.rosti@oist.jp}}}

\affiliation{Complex Fluids and Flows Unit, Okinawa Institute of Science and Technology Graduate University, 1919-1 Tancha, Onna-son, Kunigami-gun, Okinawa 904-0495, Japan}

\begin{document}
\maketitle

\begin{abstract}
This study compares turbulent channel flows over elastic walls with those over rough walls, to explore the role of the dynamic change of shape of the wall on turbulence. The comparison is made meaningful by generating rough walls from instantaneous configurations of the elastic cases. The aim of this comparison is to individually understand the role of fluid-structure interaction effects and the role of wall shape/undulations in determining the overall physics of flow near elastic walls. With an increase in the compliance of the wall, qualitatively similar trends for many of the effects produced by a rough wall are also seen in the elastic wall. However, specific features can be observed for the elastic wall cases only, arising from the mutual interaction between the solid and fluid, leading to a further increase in drag. 
To understand them, we look at the turbulent structures, which exhibit clear differences across the various configurations: roughness induces only a slight reduction of streamwise coherency, resulting in a situation qualitatively similar to what is found in classical turbulent channel flows, whereas elasticity causes the emergence of a novel dominant spanwise coherency. Additionally, we explored the effect of vertical disturbances on elastic wall dynamics by comparing with permeable walls having similar (average) wall-normal velocity fluctuations at the interface. The permeable walls were found to have minimal similarities with elastic walls. Overall, we can state that the wall motion caused by the complex fluid-structure interaction contributes significantly to the flow and must be considered when modeling it. In particular, we highlight the emergence of strong wall-normal fluctuations near the wall, which result in strong ejection events, an attribute not observed for rigid walls.
\end{abstract}



\section{Introduction}
Turbulent flows over compliant walls differ greatly from those over rigid walls due to the mutual interactions between the fluid and the wall. Understanding how elastic surfaces interact with a turbulent flow has attracted the attention of both fundamental and industrial problems, e.g., laminar-to-turbulent transition delay \citep{riley1988complaint, carpenter1993optimization, nagy2022effect}, friction drag reduction \citep{semenov1991conditions, carpenter2000hydrodynamics, gad2002compliant}, and the restraint of flow-involved noise and vibrations \citep{nisewanger1964flow}. The use of compliant surfaces to control the flow is particularly intriguing, as no additional energy is required.

Inspired by the experimental research by \citet{kramer1960boundary, kramer1962boundary} that reported drag reduction with compliant walls, theoretical studies were carried out to understand the delay in transition and turbulence attenuation; see, e.g., \cite{kumaran2021stability} for the recent review.
Classical instability analysis aimed to understand the wall compliance effect on the laminar to turbulent transition. Fluid-compliant wall interactions lead to the generation of surface waves and the modification of turbulent flows due to the deformation. \citet{benjamin1960effects, benjamin1963threefold, landahl1962stability} found that wall compliance can lead to both stabilised and de-stabilised modes. Subsequently, \citet{carpenter1985hydrodynamic, carpenter1986hydrodynamic} reported that the above instabilities can be categorised into two distinct instabilities: a Tollmien-Schlichting (TS) instability, same as observed over the rigid wall, and a flow-induced surface instability (FISI), which is suppressed by the material damping \citep{wang2006two}. TS instability is stabilized by wall compliance, resulting in the transition delay \citep[see e.g.,][]{rotenberry1990effect, rotenberry1992finite, davies1997numerical}. Conversely, for FISI, wall compliance results in two instabilities \citep{davies1997instabilities}: traveling wave flutter and static divergence. The traveling wave flutter is an advected instability generated by the phase difference between pressure fluctuations and wall-normal fluctuations of the compliant wall \citep{carpenter2000hydrodynamics}. The wave advected at 70\% of the free stream velocity, and the wave amplitude arose in the streamwise direction without temporal growth. Conversely, the static divergence wave is a relatively slow downstream traveling wave formulated along the spanwise direction with a phase speed which is 5\% of the free stream velocity \citep{gad1984interaction, lucey1992numerical}.

\citet{duncan1986response} examined the wall-interface responses by forcing it with interfacial pressure pulses, and reported relations between the coating response and the fluid speed. As the flow speed increased, the response to the pressure pulse became unstable, and at very high speeds, waves with large amplitudes rose on the surface. \citet{luhar2015framework} expresses the turbulent velocity field as a linear superposition of propagating modes, extending the resolvent framework by \citet{mckeon2010critical}. These modes were affected by the compliant wall, which the authors modeled as a complex wall admittance linking pressure and velocity. This framework predicts the emergence of the quasi-two-dimensional propagating waves observed in Direct Numerical Simulations (DNS) \citep{kim2014space}. Also, \cite{toedtli2019predicting} evaluated the capabilities of a low-order flow model based on the resolvent analysis and confirms that the attainable drag reduction in wall-bounded turbulent flows strongly depends on the relative phase between sensor measurement and actuator response. \citet{benschop2019deformation} investigated the influence of different coating parameters, such as its density, stiffness, \textit{etc.}, on the deformations of a linear visco-elastic compliant coating in a turbulent flow, using a one-way coupled model. The model was able to reproduce experiments, predicting the order of magnitude of the surface displacement and capturing its increase with Reynolds number and coating softness; also, it was found that the deformation is approximately three times larger than the wall thickness under resonant conditions and that the vertical displacements are mainly driven by the pressure fluctuations and not by the shear stress.

Experimental approaches have reported the characteristics of the interaction between surface deformation and turbulent flow, by varying the wall elasticity. When compliant walls show deformations smaller than one wall unit, \citet{zhang2015integrating} termed this regime one-way coupled, and found two types of surface motions, i.e., slow and fast traveling waves. Also, a strong correlation between turbulent pressure fluctuations and wall deformations is evident in this regime \citep{zhang2017deformation}. On the other hand, in the two-way coupled regime, when deformations are larger than several wall units, \citet{wang2020interaction} reported two modes: the spanwise-aligned waves advected in the streamwise direction at the 66\% of the free-stream velocity, and the streamwise-aligned waves advected in the spanwise direction at the material shear speed. With the increase in wall deformation, the authors also observed that the flow momentum decreases in the buffer and viscous sublayers, an observation also made by \citet{greidanus2022response}.

Early studies based on numerical simulations modeled the compliant walls as spring-mass damper systems \citep{endo2002direct,kim2014space}, where the displacement and velocity of the wall are used as time-evolving boundary conditions on the flow equations. However, these simple and easy-to-implement models usually included only the wall-normal effects coming from the pressure fluctuations to drive the motion of the wall, neglecting instead the tangential stresses. Since the latter can influence the presence of vorticity at the boundary \citep{morton1984generation} and the wave motion in an elastic layer \citep{rayleigh1885waves}, it is essential to capture the effects of viscous stresses when considering highly elastic substrates.

\citet{Rosti_Brandt_2017} accounted for the full wall motion and conducted the first DNSs of a turbulent channel flow over an incompressible viscous hyper-elastic layer, modeled as a neo-Hookean solid. They observed that the drag increases with the wall elasticity, an effect caused by the intense non-zero wall normal fluctuations owing to the elastic wall motion. They extended their discussions in light of turbulent flow over rough and porous walls by making comparisons with the results by \citet{Breugem2006The} and \citet{ORLANDI_LEONARDI_2008}. The authors also reported that the flow structures are significantly disturbed in the streamwise direction and tend to organize along the spanwise direction, with increasing wall elasticity. Due to the appearance of spanwise coherent structures, high momentum flow is brought towards the elastic wall, driving its deformation; as a consequence, the interface pushes the flow back towards the channel center, overall causing intense vertical velocity fluctuation events \citep{Ardekani_Rosti_Brandt_2019}. The strong ejections with significant wall-normal velocity, together with the small negative streamwise velocity fluctuations, contribute to the turbulent production, whereas the sweep events become stronger than ejections for high wall compliancy. Subsequently, \citet{Esteghamatian2022spatiotemporal} showed that the strong ejection events can be explained by the negative vorticity lift-up mechanism, leading to the transport of low-speed fluid away from the near-wall region. It is activated when the advection speed of near-wall pressure fluctuations matches the phase speed of Rayleigh waves, which travel along the surface of an elastic solid with their penetration depth comparable to the wavelength \citep{rayleigh1885waves}, a phenomenon caused by the inflectional velocity profile near the wave troughs. Additionally, using a dynamic surface-fitted coordinate system, they showed that compliant walls shift the logarithmic mean velocity profile downwards, with no alteration of the viscous sublayer. Recently, \citet{lu2024scaling} highlighted the role of a `critical layer' in the turbulent boundary layer over compliant walls; the critical layer was introduced by \citet{miles1957generation} to investigate the two-dimensional surface waves generated by the shear flows. They report that, below the critical layer, turbulence is phase-locked and travels with the deformation; above the layer, turbulence travels with the mean local streamwise velocity and is decoupled from the deformation. The above study is a clear example of deformation-induced modifications on turbulence.

The above literature mainly focused on studying the overall effect of compliant walls; however, the effect of elastic walls can be decoupled into a series of different effects, such as surface undulation (due to the wall deformation by the hydrodynamic force), non-zero wall-normal velocity fluctuations (originated by the dynamic wall movement), and the wall acceleration (caused by the propagation of waves on the surface and inside the materials), as done in several modeling efforts. Since the above effects appear together, it is complicated to understand which of these contributions is dominant in modifying the flow. Numerical simulations can be a useful tool to try to decouple these effects by artificially suppressing some of them, and indeed, some previous studies have made such attempts. For example, while investigating the effects of wall elasticity on a particle suspension, \citet{Ardekani_Rosti_Brandt_2019} looked at the role of roughness by performing a simulation of a turbulent channel flow over an instantaneous surface geometry taken from the fully coupled simulation, but frozen in time (thus removing the effects caused by the wall motion).
\citet{Foggi_Rota_Monti_Olivieri_Rosti_2024} adopted a similar approach while studying turbulent flows over dense canopies. The authors compared the turbulent flow over a flexible canopy to that over a `frozen' one (obtained from an instantaneous configuration of the deformable canopy), in which the filaments' motion was suppressed.

In this work, we adopt this approach to investigate the role of the wall motion of compliant walls, by separately comparing the influence of wall elasticity and the other effects. In particular, we aim to understand the two-way coupled interaction of a compliant elastic wall in a turbulent channel, by directly comparing turbulent flows over specific rough walls that share the same statistical properties of the instantaneous configuration of the elastic wall. The effect of the wall elasticity on the turbulent statistics and flow structures is investigated by drawing comparisons across the elastic and rigid cases, for different levels of the wall `complexity', i.e., the elastic shear modulus and the corresponding roughness. In general, our aim is to determine which factor, among dynamic wall movement due to the Fluid-Structure Interaction and wall undulations, plays the key role in the flow modifications seen near the elastic wall, e.g., with respect to turbulent structures; note that our focus is not to make a one-on-one comparison between elastic and other types of walls. The manuscript is organized as follows: \S~\ref{sec: formulation} describes the governing equations of the problem and the numerical procedure used to discretize it. Next, \S~\ref{sec: results of e and r} reports the results of the comparison of the turbulent statistics and the flow structures for an elastic and rough wall with a similar level of surface deformation. Appendix~\ref{asec: porous} provides a further comparison of the elastic wall with a permeable wall model, characterized by the same average wall-normal velocity fluctuations. The overall conclusions are finally summarized in \S~\ref{sec: conclusion}. Additionally, Appendix~\ref{asec: basics for continuum mechanics} provides details about the continuum mechanics of the hyperelastic materials, and Appendix~\ref{asec: ks} explains a derivation of equivalent sand grain roughness.

\section{Methodology}\label{sec: formulation}
This section first describes the governing equations and their numerical discretization in \S~\ref{subsec: governing equations} and \S~\ref{subsec: numerical procedure}, while \S~\ref{subsec: wall configuration} reports details of the wall configurations studied in this work. Here, we focus on the fully-coupled elastic wall and the corresponding rough surfaces.

\subsection{Governing equations} \label{subsec: governing equations}
The momentum conservation equation and the incompressible constraint govern the dynamics of both the fluid and elastic solid phases,
 \begin{subeqnarray}
   \frac{\partial u_i^p}{\partial t} + \frac{\partial u^p_i u^p_j}{\partial x_j} &=& \frac{1}{\rho}\frac{\partial \sigma^p_{ij}}{\partial x_j}, \\[3pt]
   \frac{\partial u^p_i}{\partial x_i}&=&0.
   \label{eq:momentum}
\end{subeqnarray}
Here, the subscript $i$ (or $j$) represents the streamwise ($x$), wall-normal ($y$), and spanwise ($z$) direction, respectively, and $u_i$ (or $u$, $v$, and $w$) are the corresponding velocity components. The suffix $p=f, s$, represents the fluid and solid phases, respectively, and $\rho$ is the density of both fluid and solid, which we assume to be the same. The fluid phase is a Newtonian fluid, while the solid is an incompressible viscous-hyperelastic material; the two have a Cauchy stress tensor $\sigma^p_{ij}$ defined as
\begin{subeqnarray}
   \sigma^f_{ij} &=& -p\delta_{ij} + 2\mu  \mathsfbi{D}_{ij}, \\[3pt]
   \sigma^s_{ij} &=& -p\delta_{ij} + 2\mu  \mathsfbi{D}_{ij} + G\mathsfbi{B}_{ij}.
   \label{eq: stress}
\end{subeqnarray}
where $p$ is the pressure and $\mu$ is the dynamic viscosity of the two phases, assumed to be the same, $\mathsfbi{D}_{ij}$ is the strain rate tensor defined by $\mathsfbi{D}_{ij} = \frac{1}{2}\left(\frac{\partial u_i}{\partial x_j} + \frac{\partial u_j}{\partial x_i}\right)$, and $\delta_{ij}$ is the Kronecker delta function. The last term in Eq.~\ref{eq: stress}$b$ is the elastic wall contribution modeled here as a neo-Hooken solid satisfying the incompressible Mooney-Rivlin law (see e.g. \citet{Bonet_Wood_2008}), where $G$ is the shear modulus and $\mathsfbi{B}_{ij}$ is the left Cauchy-Green deformation tensor. For the details about the continuum mechanics of the hyperelastic materials, see Appendix~\ref{asec: basics for continuum mechanics}. $\mathsfbi{B}_{ij}$ can be found by solving the following transport equation,
\begin{equation}
  \frac{\partial \mathsfbi{B}_{ij}}{\partial t} + \frac{\partial u_k^s \mathsfbi{B}_{ij}}{\partial x_k} = \mathsfbi{B}_{kj}\frac{\partial u_i^s}{\partial x_k} + \mathsfbi{B}_{ik}\frac{\partial u_j^s}{\partial x_k}.
  \label{eq: B transport}
\end{equation}
The solid and fluid phases are coupled at the interface by the continuity of the velocity and of the normal traction force as
\refstepcounter{equation}
$$
  u^f_i = u^s_i, \quad
  \sigma^f_{ij}n_j = \sigma^s_{ij}n_j,
  \eqno{(\theequation{\mathit{a},\mathit{b}})}
  \label{eq: continuity}
$$
where $n_j$ is the normal vector at the interface.

To solve the fluid-structure interaction problem, we apply the one-fluid formulation \citep{prosperetti2009computational} (also called one-continuum formulation) as detailed by \citet{Sugiyama2011}; in particular, we solve a single set of equations valid in both phases by employing a monolithic velocity vector field, obtained by a volume averaging procedure
\begin{equation}
  u_i = (1-\phi^s)u^f_i + \phi^s u^s_i. 
  \label{eq: one continuum formulation}
\end{equation}
Here, $\phi^s$ is the solid volume fraction, which can take values in the range $0\leq \phi^s \leq 1$, its value being zero and one in the the fluid and solid phases, and with $\phi^s = 0.5$ corresponding to the interface between the two. To close the system of equations in a purely Eulerian form, the following transport equation for $\phi^s$ is solved
 \begin{equation}
  \frac{\partial \phi^s}{\partial t} + \frac{\partial u_k \phi^s}{\partial x_k} = 0.
  \label{eq: phi transport}
 \end{equation} 

Solving Eq.~\ref{eq: B transport} directly can lead to a numerical instability due to the scattered distribution of $\mathsfbi{B}_{ij}$ in the fluid region; following \citet{Sugiyama2011}, we solve an equation for the modified left Cauchy-Green deformation tensor $\widetilde{{\mathsfbi{B}}}_{ij} = \phi_s^\alpha \mathsfbi{B}_{ij}$ (with $\alpha=1/2$) instead of Eq.~\ref{eq: B transport}. $\widetilde{\mathsfbi{B}}_{ij}$ is governed by
\begin{equation}
\frac{\partial \widetilde{\mathsfbi{B}}_{ij}}{\partial t} + \frac{\partial u_k^s \widetilde{\mathsfbi{B}}_{ij}}{\partial x_k} = \widetilde{\mathsfbi{B}}_{kj}\frac{\partial u_i^s}{\partial x_k} + \widetilde{\mathsfbi{B}}_{ik}\frac{\partial u_j^s}{\partial x_k}.
\label{eq: B transport tilde}
\end{equation}
Doing this, we have $\widetilde{\mathsfbi{B}}_{ij} = 0$ for $\phi_s^\alpha = 0$, and that the numerical instability can be avoided in the fluid region.  The initial condition of Eq.~\ref{eq: B transport} is $\mathsfbi{B}_{ij} = \mathsfbi{I}$ (unstressed material), which becomes $\widetilde{\mathsfbi{B}}_{ij} = \phi_s^\alpha \mathsfbi{I}$ for Eq.~\ref{eq: B transport tilde}.

\begin{figure}
  \centering
  \includegraphics[width=0.45\textwidth]{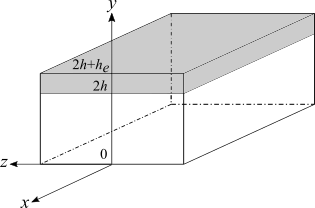}
  \caption{Sketch of the computational domain for the present set of simulations.}
  \label{fig: sketch of domain}
\end{figure}

\subsection{Numerical procedure}\label{subsec: numerical procedure}
The conservation equations are discretized in an orthogonal staggered grid using a second-order central finite difference scheme, except for the advection term in Eqs.~\ref{eq:  B transport} and \ref{eq: phi transport}, which is discretized using the fifth-order weighted essentially non-oscillatory (WENO) scheme. The discretized differential equations are advected in time using an explicit fractional-step method, where all terms are advanced by the third-order Runge-Kutta scheme, except the solid stress term in Eq.~\ref{eq:momentum}, where we use the Crank-Nicolson scheme. The numerical method has been extensively described in \citet{Sugiyama2011}, and the interested readers are directed to e.g., \citet{ii2012computational, rosti2018rheology, Rosti_Brandt_2017, alghalibi2019inertial, rosti2020breakdown} for more details and validations.

Figure~\ref{fig: sketch of domain} illustrates the computational domain and the chosen Cartesian coordinate system. The domain is a box bounded at the bottom ($y=0$) and at the top ($y = 2h + h_e$) by two solid impermeable walls; the channel is occupied at the top by the elastic wall of height $h_e$, and the rest $2h$ by the fluid. A no-slip and no-penetration boundary is imposed on the (top and bottom) rigid wall, and periodic boundary conditions are enforced in the streamwise and spanwise directions. For all simulations, a uniform grid is employed with $1296 \times 540 \times 648$ grid points in a domain of size $6h \times 2.5h \times 3h$ in the streamwise, wall-normal, and spanwise directions, respectively. The spatial resolution is such that $\Delta x^{w+} = \Delta y^{w+} = \Delta z^{w+} < 0.8$, and are the same used by \citet{Rosti_Brandt_2017, Ardekani_Rosti_Brandt_2019, rosti2020low}, who also ensured the grid independence of the results. This very fine and cubic, uniform grid is chosen to properly resolve Eqs.~\ref{eq: B transport} and \ref{eq: continuity}. Eq.~\ref{eq: B transport} does not have an explicit diffusion term, which may pose numerical difficulties; to address this issue, a common approach is to introduce artificial dissipation, or, alternatively, to use specialised numerical schemes, such as a third-order compact upwind scheme, often used in the past for polymeric turbulent flows (who shares a similar equation and problem, see e.g., \citet{dubief2005new, dupret1986loss,min2001effect}), or a fifth-order weighted essentially non-oscillatory (WENO) scheme. 
For the same accuracy, the WENO scheme is relatively less expensive computationally; \citet{izbassarov2018computational} investigated the performance of the WENO scheme in a polymeric flow, and showed that any need for local artificial diffusion can be removed using smaller grids, which is why in the present study we use the WENO scheme with a fine grid. This method was also used before by \citet{ Ardekani_Rosti_Brandt_2019, Rosti_Brandt_2017} for turbulent flows over elastic walls. Note that, the effect of implicit numerical diffusion is limited by the high order of the WENO scheme chosen (5th order), and its effect was tested through a grid refinement study by \citet{Rosti_Brandt_2017}. 
The temporal resolution is chosen such that the Courant-Friedrichs-Lewy (CFL) number is equal $0.25$.

Note that the values with superscripts $()^{w+}$ and $()^+$ represent the values normalized by the wall units (either the friction velocity or the viscous length scale), corresponding to the bottom rigid wall and the complex top wall, respectively. In particular, the friction velocity of the bottom rigid wall $u_\tau^{w}$ can be obtained as usual as
\begin{equation}
  u_\tau^w = \sqrt{\mu \frac{\mathrm{d}\bar u}{\mathrm{d}y} \Bigg|_{y=0}},
  \label{eq: rigid wall friction velocity}
\end{equation}
where $\bar{u}$ represents the Reynolds averaged value of $u$, accompanied with $u^\prime$ that represents the deviation of $u$ from the mean, i.e., $u = \bar{u} + u^\prime$. The friction velocity $u_\tau$ at the complex wall is instead defined as
\begin{equation}
  u_\tau = \sqrt{\mu\frac{\mathrm{d}\bar u}{\mathrm{d}y} \Bigg|_{y=2} - \rho \overline{u^\prime v^\prime}|_{y=2} + G \overline{\mathsfbi{B}}_{xy}|_{y=2}}.
  \label{eq: complex wall friction velocity}
\end{equation}
The values of $u_\tau^w$ and $u_\tau$ are related to the balance between the driving streamwise pressure gradient and the total wall shear stress. Indeed, a constant flow rate condition is enforced to fix the bulk velocity $U_b$ (the average mean fluid velocity across the whole domain) by adapting the streamwise pressure gradient at every time step.

The above numerical methods and details are kept the same for both the fully coupled elastic simulation and for the case with the rigid rough wall. However, since we do not need to solve for the solid velocity and stress in the latter, we thus rely on the simpler immersed boundary method based on the volume penalization \citep{kajishima2001turbulence, schneider2005numerical, yuki2007efficient, breugem2014flows, Ardekani_Rosti_Brandt_2019, kumar2024implementation}.

\subsection{Details of the wall  surface}\label{subsec: wall configuration}
In this section, we provide more details on how the elastic and rough walls are configured in this study.

\begin{figure}
  \centering
  \includegraphics[width=1\textwidth]{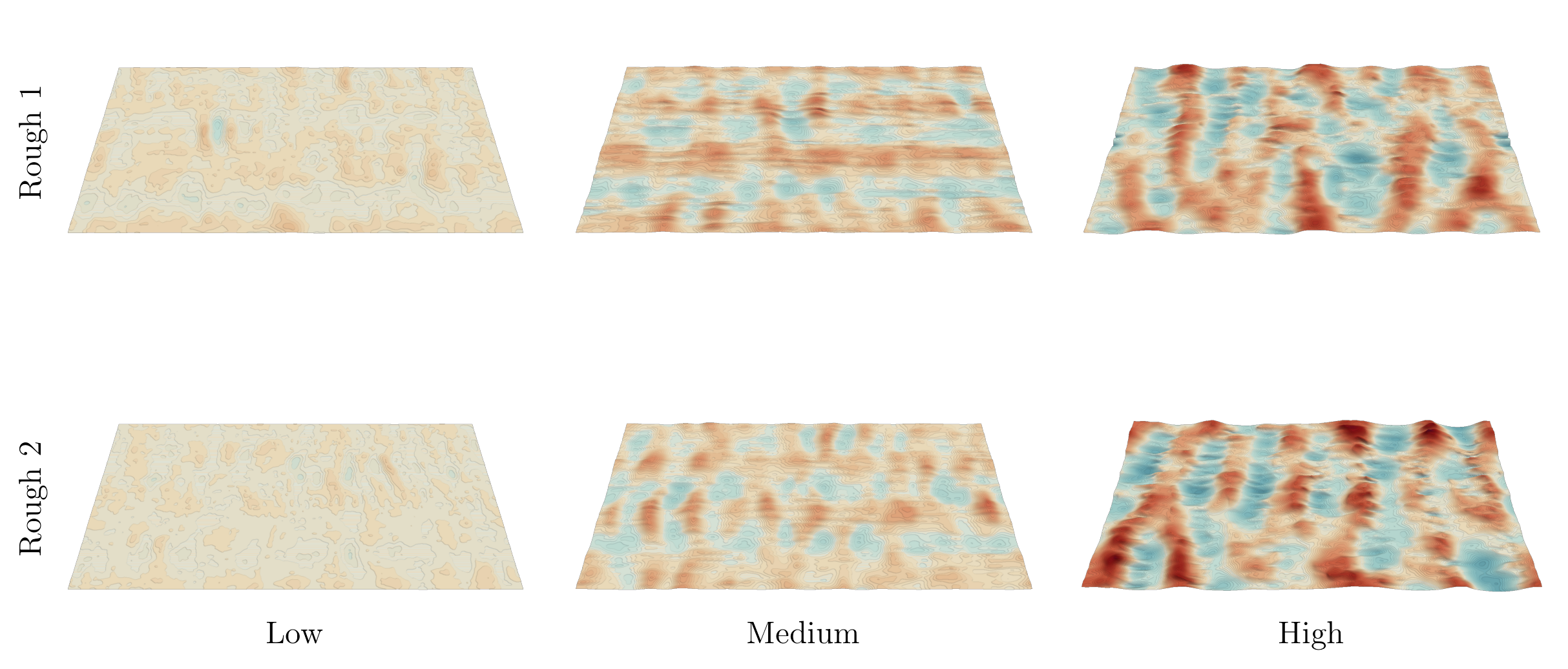}
  \caption{Instantaneous configurations of the wall for the rough cases. The two rows represent the rough 1 (top) and rough 2 (bottom) cases, and the columns correspond to an increasing level of deformation, from low (left) to high (right). The color shows the vertical displacement of the wall $\delta$, ranging from $-0.2h$ (blue) to $0.2h$ (red), with negative and positive values corresponding to displacements towards the solid and fluid regions, respectively.}
  \label{fig: wall shape}
\end{figure} 

In this study, the viscosity ratio between solid to fluid phases, $\mu^s/\mu^f =1$. The work of \citet{Rosti_Brandt_2017} investigated the effect of the viscosity ratio $\mu^s/\mu^f$ on the turbulent dynamics. For a fixed value of $G$, they saw that statistics is mainly governed by the flow dynamics when the ratio is smaller than 1. However, when it is larger than 1, the behavior, especially at the interface, is determined by the solid properties themselves. We chose the value of 1 to strike a balance between the fluid and solid contributions. In this study, the wall elasticity behavior is described by the hyperelastic material (large deformations, nonlinear, strain-dependent), which also has a viscous component attached to it (strain-rate dependence), e.g., gel-like substances (\citet{verma2013multifold}), or biological tissues (\citet{sugiyama2010full}). For practical applications such as tubes (such as blood vessels) bounded by solids such as polymeric gels, \citet{kumaran1995stability} discusses solid and fluid viscosities of similar magnitudes, i.e., with viscosity ratios $O(1)$.

The interface between the elastic wall and the fluid is initially flat and parallel to the rigid walls in unstressed conditions; following previous studies, we fix the elastic wall thickness to be equal to $h_e=0.5h$ \citep{Rosti_Brandt_2017}.  The control parameter used to characterize the elasticity of the wall is the modulus of transverse elasticity, $G$, which is varied from $G/(\rho U_b^2)=2.0$ (almost rigid wall) to $G/(\rho U_b^2)=0.5$ (very deformable/highly elastic wall). After attaining a fully developed turbulent flow, an instantaneous surface configuration of the elastic wall is taken and used as a rough surface in a different simulation. Effectively, the roughness is controlled by the magnitude of the deformation of the walls (that is, the higher the deformation, the higher is the roughness), which is associated with the shear modulus of the original elastic wall case. Numerically, the roughness effect of the wall is maintained by not updating the transport equations for $\mathsfbi{B}_{ij}$ (Eq.~\ref{eq: B transport}) and $\phi^s$ (Eq.~\ref{eq: phi transport}), thus suppressing the fluid-solid interaction. In the present study, we generate two different rough wall configurations from each level of elasticity, denoted as rough 1 and 2, to ensure the generality of our results. This is done by choosing two different instantaneous interface geometries of the elastic wall at different time instants.

\begin{figure}
  \centering
  \includegraphics[width=1\textwidth]{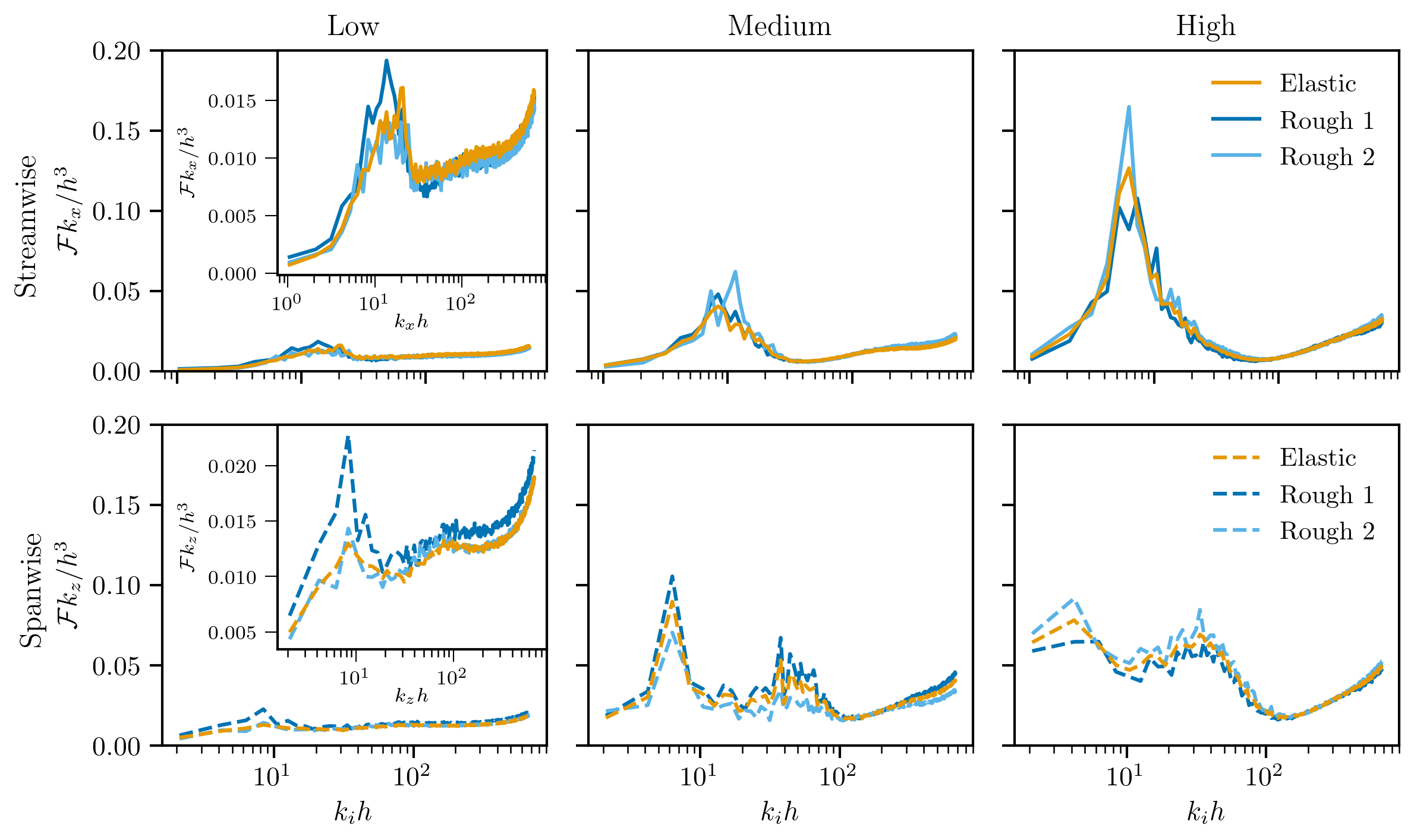}
  \caption{One-dimensional premultiplied power spectrum $\mathcal{F}k_i$ of the wall deformation, as a function of the wavenumber. The top and bottom rows show the results for the streamwise (top) and spanwise (bottom) direction, and the columns represent the level of wall elasticity/surface deformation: low (left) to high (right). The line style represents the directions of the analysis, streamwise (solid line) and spanwise (dashed line), and the line color shows the different cases: elastic (orange), rough 1 (dark blue), and rough 2 (light blue). The inserted figures in the top-left and bottom-left show enlarged views.}
  \label{fig: psd}
\end{figure}

In Figure~\ref{fig: wall shape}, we show the surface visualizations for the two rough cases analyzed in this study. Some characteristic features can be assessed from the figure: the case with a low level of roughness exhibits a multitude of small-scale features, whereas that with large roughness shows a characteristic wavelength in the streamwise direction, of about $h$, while a high level of coherency is visible in the spanwise direction. To fully characterize the undulations of the walls, we calculate the spectrum $\mathcal{F}$ in the streamwise and spanwise directions of the surface displacement $\delta$, defined as the roughness height from the reference flat surface at $y = 2h$, i.e., $\delta(\boldsymbol{x}) = y(\boldsymbol{x}) - 2.0h$. Figure~\ref{fig: psd} shows these one-dimensional premultiplied power spectra, $\mathcal{F}k_i$ ($i$ representing the $x$ or $z$ directions), for different degrees of deformation going from low (left) to high (right). The top panel shows the spectrum against the streamwise direction and the bottom panel for the spanwise; the line color distinguishes the different cases: elastic (orange), rough 1 (dark blue), and rough 2 (light blue). The distributions demonstrate the similarity of the different wall configurations analyzed in this work at each level of deformation. Consistent with the previous assessments, Figure~\ref{fig: psd} (top-left to right) shows that a clear dominant peak emerges in the streamwise wavelength with an increase in the wall elasticity. In the low and intermediate case, there are structures repeating in both span and stream directions, and a visible peak is observed in both directions in the spectra. Subsequently, as the degree of deformation increases, the spanwise peak comes down, and the streamwise peak becomes dominant, see Figure~\ref{fig: psd} (the rightmost column), due to the formation of rollers aligned along the span, as will be discussed later. Overall, this analysis demonstrates the similarities between the rough and instantaneous elastic wall configuration, such that we can now compare how the flow reacts to it in the following sections.

\begin{table}
  \begin{center}
  \def~{\hphantom{0}}
  \begin{tabular}{c|c|cc|ccccc}
        {Case}&  {Deformation}&  {$G/(\rho U_b^2)$}&  {$G/(\rho u_\tau^2)$}& 
        {$Re_\tau^w$}&  {$Re_\tau$}&  {$\delta_{max}/h$}&  {$\delta_{rms}/h$}&  {$\mathcal{S}$}\\[3pt]
       \hline
        {Smooth}& {--}& {$\infty$}& { $\infty$}& {178.5       }& { 178.5}& {0}& {0}& {0}\\ 
       \hline
        {Elastic}& {Low}& {2.0}& {410.9}& {183.0}& {195.4}& {0.027}& {0.007}& {0.286}\\
                  & {Medium}& {1.0}& {136.7}& {192.1}& {239.5}& {0.079}& {0.021}& {0.489}\\
                  & {High}& {0.5}& {33.7}& {212.0}& {341.0}& {0.180}& {0.042}& {0.649}\\[2pt]
       \hline
        {Rough}& {Low}& {$\infty$}& {$\infty$}& {178.8}& {184.7}& {0.026}& {0.007}& {0.267}\\
                  & {Medium}& {$\infty$}& {$\infty$}& {179.6}& {212.8}& {0.079}& {0.019}& {0.463}\\
                  & {High}& {$\infty$}& {$\infty$}& {189.4}& {314.2}& {0.154}& {0.037}& {0.619}\\
  \end{tabular}
  \caption{Summary of the cases investigated in this work. The table reports the shear modulus normalized by the bulk quantities $G/(\rho U_b^2)$, the shear modulus normalized by the inner scale $G/(\rho u_\tau^2)$, the rigid wall friction Reynolds number $Re_\tau^w$, the complex wall friction Reynolds number $Re_\tau$, the maximum $\delta_{max}/h$ and root mean square $\delta_{rms}/h$ values of the wall deformation, and the root mean square of the surface slope $\mathcal{S}$.}
  \label{tab: summary wall config and flow chara}
  \end{center}
\end{table}

\section{The effect of the wall roughness}\label{sec: results of e and r}
All simulations are performed at a constant flow rate, providing a bulk Reynolds number $Re = \rho U_b h/ \mu = 2800$. The full set of the simulations is reported in Table~\ref{tab: summary wall config and flow chara}, which also reports the resulting friction Reynolds number based on the bottom rigid wall $Re_\tau^{w} = u_\tau^{w} h/\nu$ and on the top complex wall $Re_\tau = u_\tau h/\nu$, as well as the maximum $\delta_{max}/h$ and root mean square $\delta_{rms}/h$ displacement of the wall, and the root mean square of the slope of the surface $\mathcal{S}$, defined as
\begin{equation}
  \mathcal{S} = \sqrt{\dfrac{1}{A} \displaystyle \int_A \left(\dfrac{\partial \delta}{\partial x}\right)^2 + \left(\dfrac{\partial \delta}{\partial y}\right)^2 ~dA},
  \label{eq: Sdq}
\end{equation}
where $A$ is the surface area. Note that, from here onward, we have ensemble averaged the results from the two rough cases (rough 1 and 2), and we show them as a single case, addressed simply as rough. The table shows that the friction Reynolds number of both walls increases with the wall deformation, especially for the elastic walls, with a value around $180$ at low elasticity and $341$ for the high elasticity. A similar trend is observed for the rough wall, although the levels achieved are always smaller than in the elastic case. Figure~\ref{fig: Re_tau} shows a visual representation of this trend, expressed in terms of the percentage of drag increase $DI$ against the level of deformation, where the drag change is computed as
\begin{equation}
    DI [\%] = \frac{C_{f,\ Complex}^2 - C_{f,\ Smooth}^2}{C_{f,\ Smooth}^2} \times 100.
\end{equation}
Here, $C_f$ is the skin friction coefficient defined as $C_f = 2 \tau_w / \rho U_b^2$, where $\tau_w = \rho u_\tau^2$ is the wall shear stress. For low deformations, both elastic and rough walls tend to converge to the value attained for a rigid flat wall, thus with $DI=0$, while the drag increases monotonically when the deformation grows. This observation indicates that the shape of the wall, similar in the two configurations, contributes significantly to the increased drag, with the movement of the elastic wall bringing additional effects that result in a further enhancement of the drag.

\subsection{Turbulent statistics}
In this section, we draw comparisons between the elastic and rough walls by comparing their turbulent statistics. 
\subsubsection{Mean velocity}
\begin{figure}
  \centering
  \includegraphics[width=0.5\textwidth]{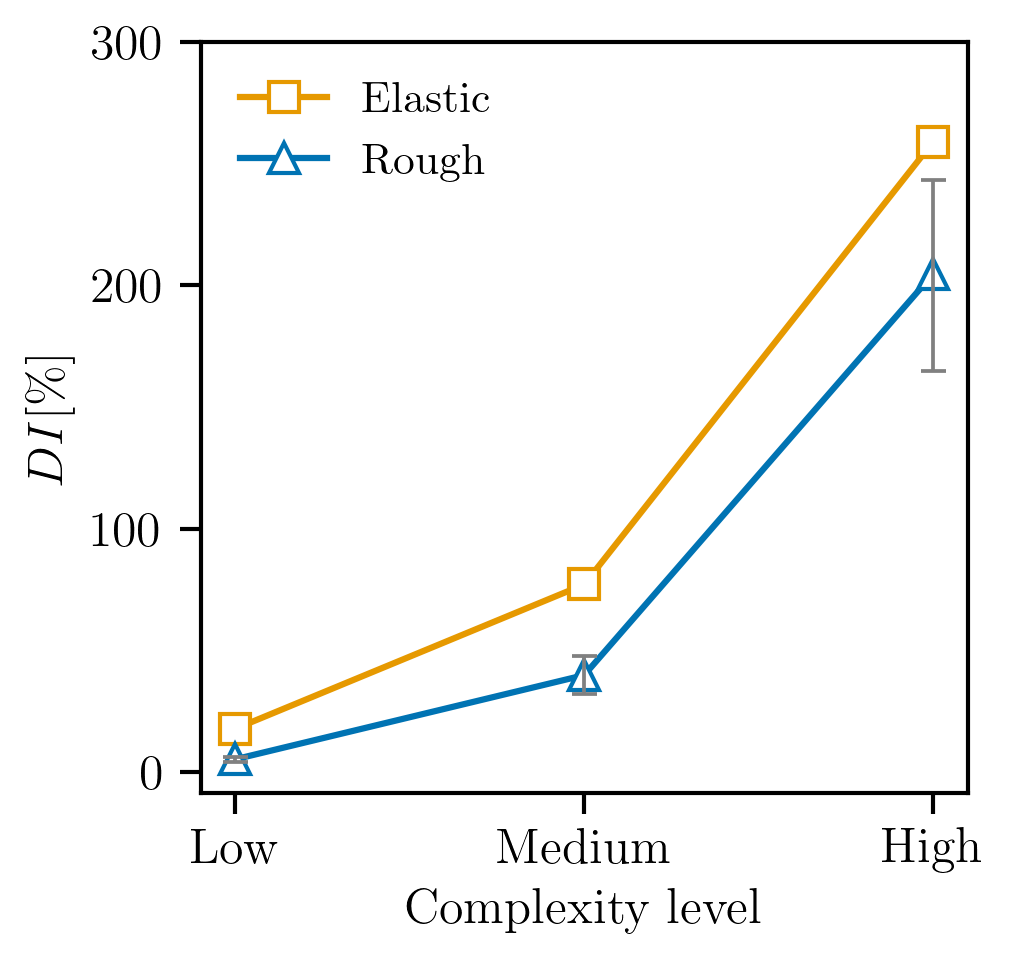}
  \caption{Percentage of drag increase $DI$ of the complex wall as a function of the level of deformation. The grey bars indicate the range of the values obtained independently from the two rough walls.}
\label{fig: Re_tau}
\end{figure}

\begin{figure}
  \includegraphics[width=1\textwidth]{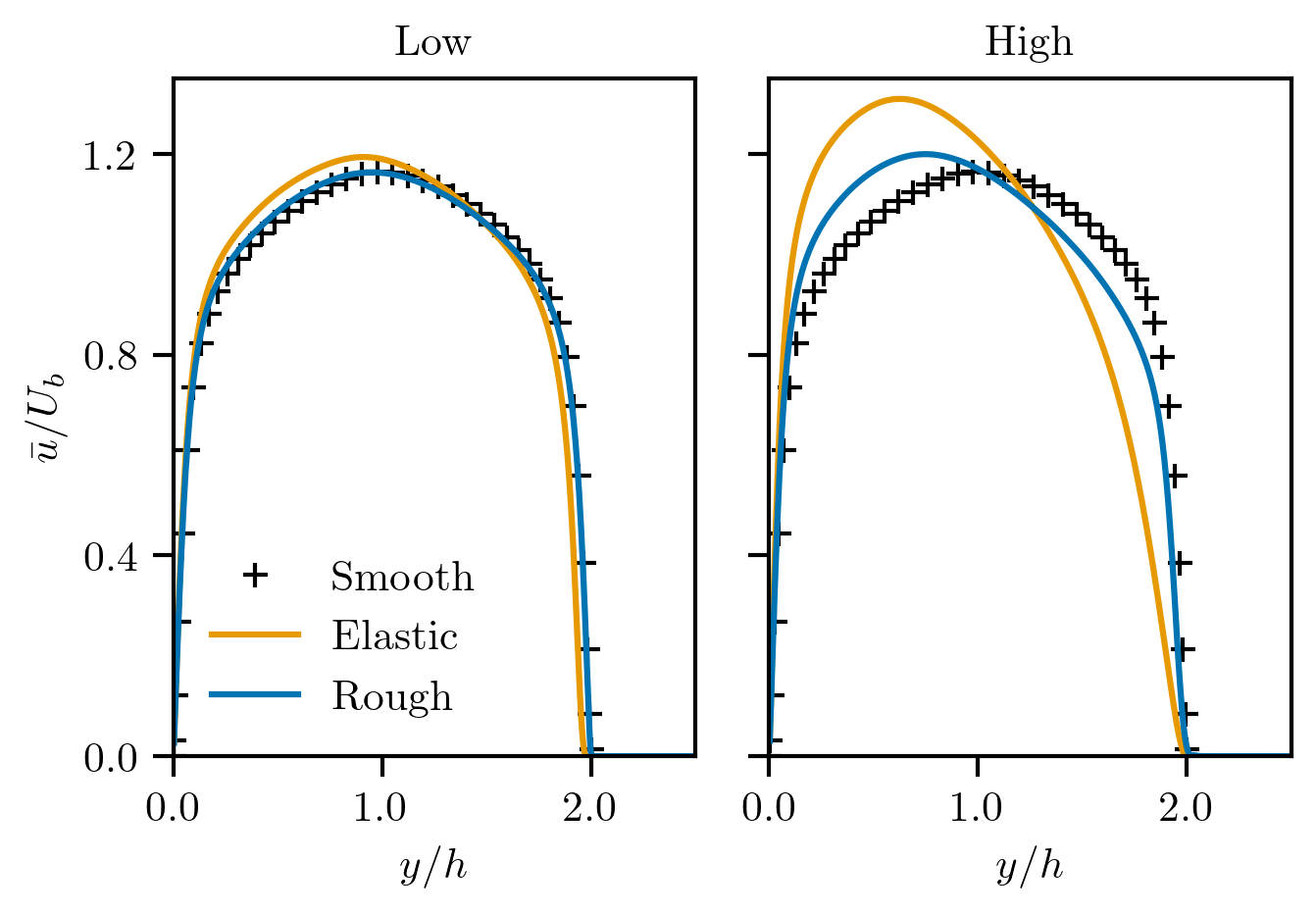}
  \caption[]{Streamwise mean velocity profiles of the elastic (orange line) and rough (blue line) walls for (left) low and (right) high levels of deformation. The $+$ symbols represent the results with smooth and rigid walls, taken from \citet{kim1987turbulence}.}
  \label{fig: MeanVel}
\end{figure}

To investigate the effect of the complex walls on the velocity field, the mean streamwise velocity profiles of the elastic and rough cases are plotted in Figure~\ref{fig: MeanVel}, and compared with those from a classical planar channel flow with smooth rigid walls \citep{kim1987turbulence}, which is thus symmetric by construction. The two panels show the cases with low (left) and high (right) deformation of the wall, while the symbols and colors correspond to the type of wall: $+$ is used for the smooth rigid wall, orange line for the elastic wall, and blue for the rough wall. It can be seen that both the elastic and rough wall profiles are skewed towards the left, i.e., the side with the smooth wall, with the effects being especially prominent for the elastic wall. As a consequence of the skewed velocity profile, the location of the maximum velocity is not in the center of the fluid region ($y=h$), but it's located closer to the bottom rigid wall. Also, the maximum velocity increases, only slightly for the rough wall, but significantly when the wall is elastic. The increased maximum velocity is accompanied by a reduction of the gradient of the velocity profile close to the elastic wall, thus indicating a reduced viscous stress contribution, notwithstanding the larger value of the friction Reynolds number and drag (see Table~\ref{tab: summary wall config and flow chara} and Figure~\ref{fig: Re_tau}). Finally, the mean velocity is null inside the solid for all walls, being rigid or elastic. These observations taken from the mean velocity profile already show significant differences between the two configurations, suggesting that the two-way fluid-solid interactions have a major contribution to the overall flow dynamics, and are able to alter the flow more significantly and globally than rigid alterations of the wall shape.

\begin{figure}
  \includegraphics[width=1\textwidth]{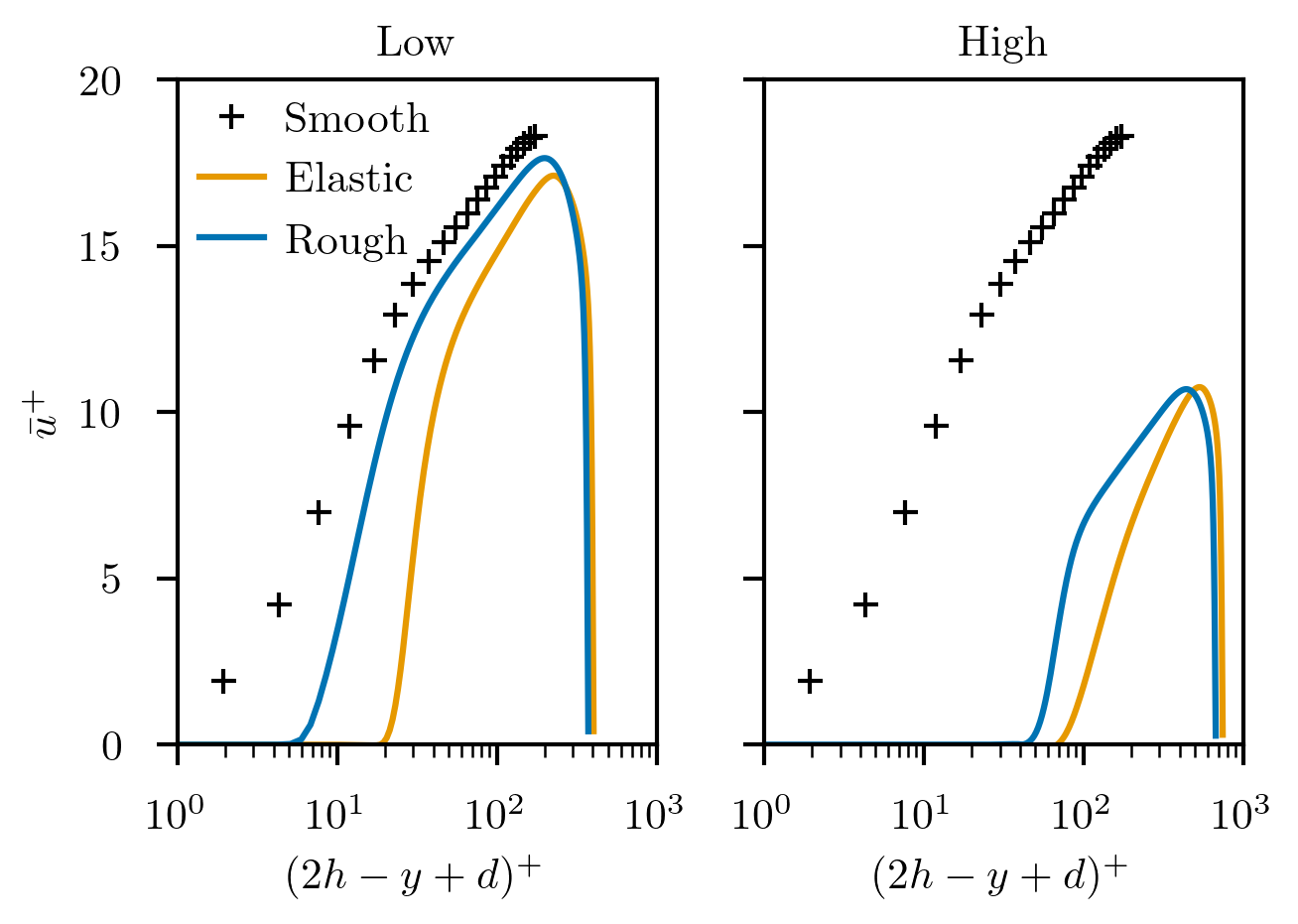}
  \caption[]{Streamwise mean velocity profiles in wall units from the different walls, for the cases with low (left) and high (right) deformations. The symbols and line colors are the same as in Figure~\ref{fig: MeanVel}.}
  \label{fig: CWallMeanVel}
\end{figure}

Figure~\ref{fig: CWallMeanVel} shows the mean velocity profiles expressed in terms of the wall units. In the logarithmic region, we follow the modified log-law
 \begin{equation}
  \Bar{u}^{+} = \frac{1}{k + \Delta k} \log (\Tilde{y}+d)^+ + B - \Delta U^+, 
 \label{eq: fitting}
 \end{equation}
where the coefficients $k$ is the von Kármán constant, $\Tilde{y}=2h-y$ is the wall-normal distance from the location of the complex walls, $d/h$ is a displacement of the origin used to take into account the wall undulation, $B$ a vertical shift, and $\Delta U^+$ is the downward shift of the velocity from the smooth channel case. While the values $k$ and $B$ for an impermeable smooth wall at moderate Reynolds number are the classical ones, i.e., $k=0.40$ and $B=5.5$, the remaining parameters ($d/h$, $\Delta k$, and $\Delta U^+$) can be found by fitting Eq.~\ref{eq: fitting} in the logarithmic region. More details on the procedure can be found in \citet{Rosti_Brandt_2017}, and the obtained values are reported in Table~\ref{tab: summary fitting}. Note that the type of fitting method could lead to potentially different values of the von Kármán constant. \cite{chen2023examination} reported that the fitting method can be affected by Reynolds number and dense roughness. However, our interests lie not in the absolute value of the von Kármán constant but in its relative variation due to the complex walls.

\begin{table}
  \begin{center}
  \def~{\hphantom{0}}
    \begin{tabular}{ccc|clccc|clccc}
                                 &                                  &         & \multicolumn{5}{c|}{Modified von Kármán constant}     & \multicolumn{5}{c}{Forcing outer-layer similarity}      \\ \hline
    \multicolumn{1}{c|}{Wall}    & \multicolumn{1}{c|}{Deformation} & $y_M/h$ & $d/h$ & $d^+$ & $k+\Delta k$ & $\Delta U^+$ & $k_s^+$ & $d/y_M$ & $d^+$ & $k+\Delta k$ & $\Delta U^+$ & $k_s^+$ \\ \hline
    \multicolumn{1}{c|}{Elastic} & \multicolumn{1}{c|}{Low}         & 1.09    & 0.07  & 13.7  & 0.29         & 6.7          & 6.5     & 0.0     & 0.0   & 0.4          & 4.0          & 16.4    \\
    \multicolumn{1}{c|}{}        & \multicolumn{1}{c|}{Medium}      & 1.23    & 0.13  & 31.1  & 0.22         & 16.4         & 42.5    & 0.05    & 19.6  & 0.4          & 11.3         & 152.4   \\
    \multicolumn{1}{c|}{}        & \multicolumn{1}{c|}{High}        & 1.39    & 0.19  & 64.8  & 0.20         & 24.6         & 308.3   & 0.09    & 54.2  & 0.4          & 15.3         & 755.0   \\ \hline
    \multicolumn{1}{c|}{Rough}   & \multicolumn{1}{c|}{Low}         & 1.05    & 0.03  & 5.54  & 0.37         & 1.9          & 2.8     & 0.0    & 0.0   & 0.4          & 0.3         & 3.7     \\
    \multicolumn{1}{c|}{}        & \multicolumn{1}{c|}{Medium}      & 1.11    & 0.09  & 19.2  & 0.36         & 5.2          & 8.9     & 0.04    & 9.9   & 0.4          & 5.7          & 15.0    \\
    \multicolumn{1}{c|}{}        & \multicolumn{1}{c|}{High}        & 1.25    & 0.15  & 47.1  & 0.36         & 11.5         & 93.8    & 0.08    & 26.9  & 0.4          & 10.0         & 90.6   
    \end{tabular}
  \caption{Summary of the coefficients of the log-law with different fitting methods. In particular, the table reports the values of the peak position of the mean streamwise velocity $y_M$, the wall-normal shift of the origin $d/h$ and $d^+$ or $d/y_M$, the modified von Kármán constant $k+\Delta k$, the logarithmic shift $\Delta U^+$, and the equivalent sand grain roughness $k_s^+$.}
  \label{tab: summary fitting}
  \end{center}   
\end{table}

When increasing the level of deformation of the wall in both the elastic and rough cases, all velocity profiles shift rightwards and downwards, with the former effect being directly caused by a progressive increase of $d/h$ due to the enhanced wall undulations, and with the latter providing a positive value of $\Delta U^+$, often associated to drag increase. The peculiar effect of the wall elasticity compared to the rough case is dual: first, the rightwards and downward shifts for the elastic cases are larger than that of the corresponding rough ones, with values of $d/h$ and $\Delta U^+$ more than double; furthermore, when the wall is elastic, the slope of the inertial range is modified.
The rightward shift of the mean velocity profile is a feature already introduced in the literature when studying turbulent flows over rough walls \citep[see e.g.,][]{Jackson_1981}, porous media \citep[see e.g.,][]{Breugem2006The, kuwata_lattice_2016, chu2021transport}, compliant walls \citep{Rosti_Brandt_2017, Ardekani_Rosti_Brandt_2019}, and canopy flow \citep[see e.g.,][]{poggi2004effect, nepf2008flow, Monti2019Largeeddy}. Instead, the change in slope of the logarithmic region, and thus the change of the von Kármán constant, has been less reported, except for some recent numerical \citep{Rosti_Brandt_2017, Ardekani_Rosti_Brandt_2019} and experimental \citep{wang2020interaction,greidanus2022response} works with elastic walls. For turbulent flows over permeable and rough walls instead, while some authors reported slight changes of the slope \citep{Breugem2006The, kuwata_suga_2017_direct, kuwata_lattice_2016}, others reported no slope variation \citep{Perry_Lim_Henbest_1987, bhaganagar2004effect, jimenez2004turbulent, flack2007examination, hong2011near, Ma_Xu_Sung_Huang_2020, Womack_Volino_Meneveau_Schultz_2022}, with the discrepancy suggested to be caused by the limited Reynolds number considered \citep{chen2023examination}. Our results in the rough case agree with the group that does not show the slope variation, suggesting that the change of slope is a peculiar feature of the elastic cases only.

We can also obtain the equivalent sand grain roughness $k_s^+$, as
\begin{equation}
    \ln k_s^+ = \ln \Tilde{y} - \frac{\kappa}{k+\Delta k}\ln (\Tilde{y}+d) + \kappa(8.5-B+\Delta U^+),
    \label{eq: my ks b}
\end{equation}
with the obtained values listed in Table \ref{tab: summary fitting}. See Appendix~\ref{asec: ks} for the derivation of Eq.~\ref{eq: my ks b}. Since $k_s^+$ is a function of the wall-normal direction, we take the values at $\Tilde{y}^+=d^+$ as representative $k_s^+$. The medium and high elastic cases, as well as the high roughness case, can be considered to be in a fully rough regime, following Nikuradse's criteria for the fully rough regime: $k_s^+>70$ \citep{pope2001turbulent}. In general, $k_s^+$ becomes larger with an increase in the wall elasticity/deformation, showing a growth comparable to the $DI$ (Figure~\ref{fig: Re_tau}). 
The values of $k_s^+$ are quite large for the highly elastic and rough cases; to assess the impact of the fitting method on our results, we also applied the method introduced by~\cite{chen2023examination}, which enforces the outer-layer similarity. The observed trends are similar to that reported above, but the numerical values of $k_s^+$ are consistently larger, thus hinting towards the fact that the variation of the von Kármán constant may not be an artifact of the previous fitting procedure but a necessary feature for these flows. However, in the present study, we do not focus on the absolute value of the fitting parameters themselves, but rather on their trends with respect to the variation of the roughness/elasticity, which remain consistent among the different methods tested; both methods seem to provide values of $k_s^+$ very large, thus further investigations regarding the optimal fitting method and the examination of the outer layer similarity in turbulent flows over elastic walls are needed.

\begin{figure}
  \centering
  \includegraphics[width=1\textwidth]{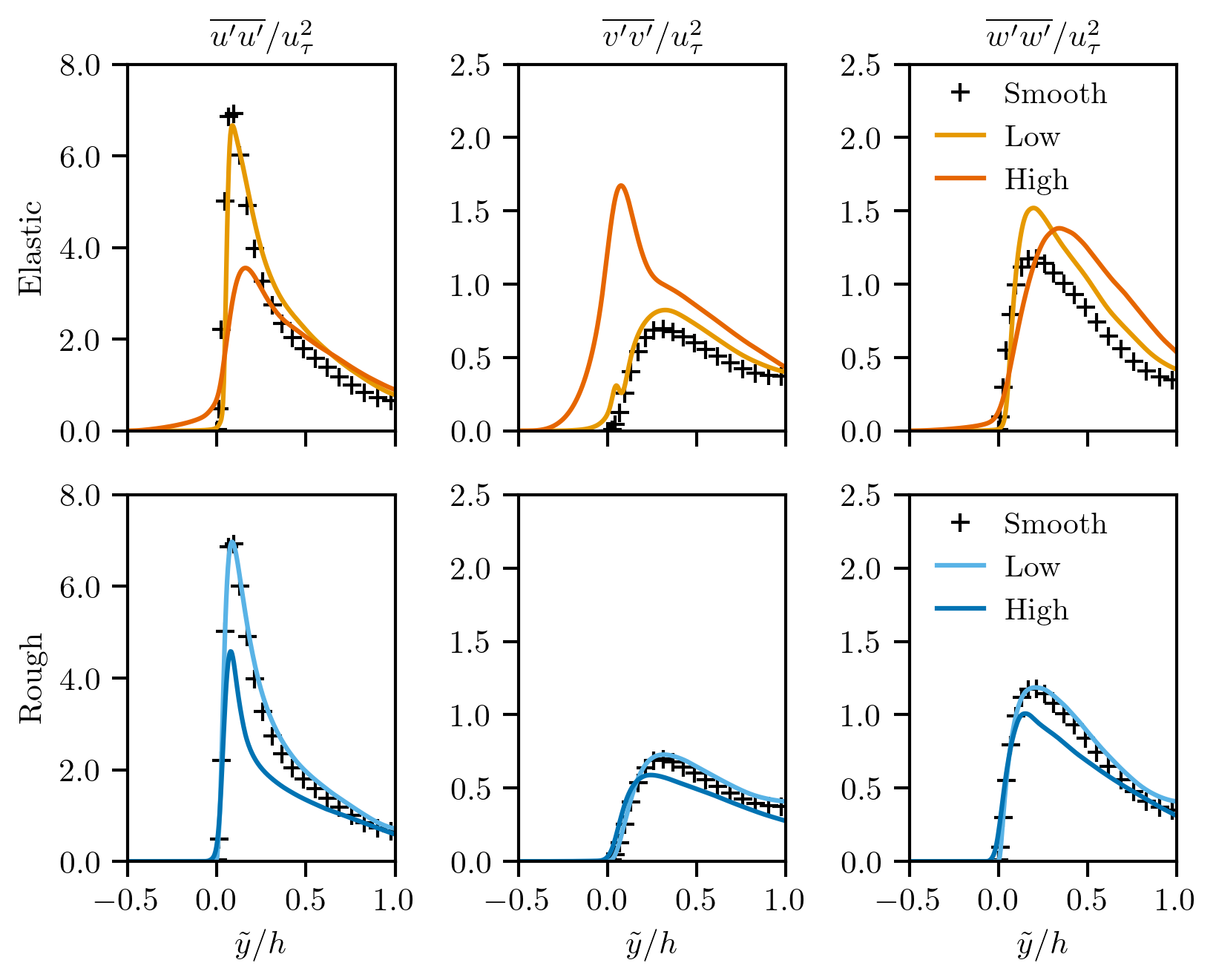}
  \caption[]{Reynolds stress components normalized by the friction velocity near the complex wall $u_\tau$, along the wall-normal direction for low (light) and high (dark) wall deformations in the (top) elastic and (bottom) rough cases. The three columns represent the different diagonal components of the Reynolds stress tensor: (left) streamwise, (middle) wall-normal, and (right) spanwise components. The symbols $+$ represent the case of a rigid, smooth wall from \cite{kim1987turbulence}.}
  \label{fig: RSS}
\end{figure}

\begin{table}
  \begin{center}
  \def~{\hphantom{0}}
  \begin{tabular}{c|c|ccccc}
      Case & Deformation &  $\overline{u^\prime u^\prime}/u_\tau^2$ &  $\overline{v^\prime v^\prime}/u_\tau^2$ &  $\overline{w^\prime w^\prime}/u_\tau^2$  \\[3pt]
      \hline
       Smooth &  -- &  0.086 &  0.29~ &  0.20~  \\[3pt]
      \hline
       Elastic &  Low &  0.086 &  0.32~ &  0.20~  \\
                 &  High &  0.16~ &  0.076 &  0.34~  \\[2pt]
      \hline
       Rough &  Low &  0.086 &  0.32~  &  0.21~  \\
                 &  High &  0.076 &  0.24~  &  0.15~  \\
  \end{tabular}
  \caption{Summary of the location of peak values of the diagonal Reynolds stress components for the low and high cases in outer scale. The location was measured from the average complex wall interface, $\Tilde{y}=0$. The data of the smooth wall was taken from \cite{kim1987turbulence}.}
  \label{tab: summary peak rss}
  \end{center}
\end{table}

\subsubsection{Reynolds shear stress}
Figure~\ref{fig: RSS} shows the profiles of the diagonal components of the Reynolds stress tensor for the elastic and rough walls, together with the classical planar channel case as a reference. The Reynolds stresses are normalized here by the friction velocity near the complex wall, $u_\tau$. In the classical channel flow, the distributions of the Reynolds stresses are symmetric, with a prevalent streamwise component, and all the components show peaks near the walls. Similarly to what was observed for the mean velocity profile, all Reynolds stress components exhibit asymmetrical distributions for the complex walls (only the wall-normal range close to the complex wall is shown here). All the Reynolds stress components are affected by the non-smooth walls, clearly demonstrating the variation with respect to wall deformation. Two major differences are evident between the elastic and rigid cases. Firstly, for the elastic wall, the components of the Reynolds stress tensor most affected are the wall-normal and spanwise ones, which significantly increase for the case with high elasticity. The streamwise component dropping, along with the spanwise and wall-normal components reinforcing, is often associated with the reduction (or absence) of the streamwise streaks, as will be discussed later in \S~\ref{sec: turbulent structure of e and r}. Also, the strong modifications of the wall-normal velocity fluctuations in the compliant wall case are due to the weakening of the wall blocking and wall-induced viscous effects \citep{Perot_Moin_1995_part1, Perot_Moin_1995_part2}, since the elastic wall can deform and better adapt to the fluid motion \citep{Ardekani_Rosti_Brandt_2019}. Secondly, the effect of the elastic wall propagates significantly inside the fluid region, with the location of the peak of the Reynolds stresses shifting away from the average wall interface location ($\Tilde{y}=0$) with an increase in the wall elasticity. This is more pronounced for the stream and spanwise components and is confirmed from Table~\ref{tab: summary peak rss}, which summarizes the wall-normal location (in outer scale) of the peak values of velocity fluctuations.
These two effects are not present for the rough wall, in which all the Reynolds stress components decrease almost uniformly with the variation of roughness, effectively maintaining the relative weights among the various components similar to that of the canonical turbulent channel flows, and thus the streamwise component is still predominant compared to the others. Also, while the distributions are asymmetric, the effect of the rough wall does not propagate much into the fluid region, with the location of the maximum fluctuations always remaining close to the wall, independently of the level of roughness, as shown in Table~\ref{tab: summary peak rss}. These roughness effects are consistent with previous experiments \citep{hong2011near, Talapatra_Katz_2012} and simulations \citep{ikeda2007direct, PIOMELLI2019}. 

\begin{figure}
  \centering
  \includegraphics[width=1\textwidth]{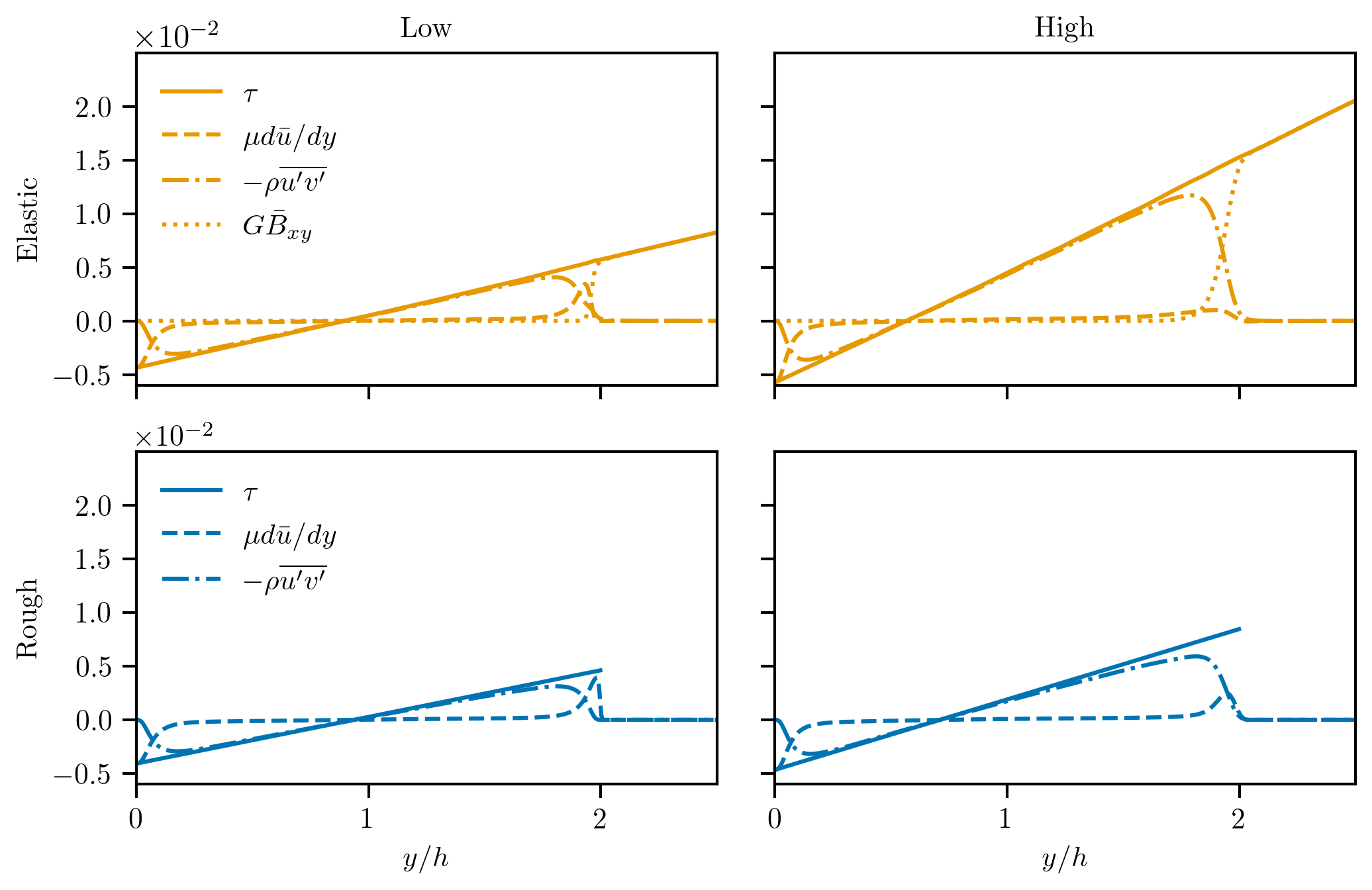}
  \caption[]{Profile of the components of the shear momentum budget for the cases with low (left) and high (right) wall deformation, for the elastic (top, orange) and rough (bottom, blue) walls. The line style represents the different stress components.}
  \label{fig: TotalRSS}
\end{figure}

We conclude this section on turbulent statistics by discussing the budget of the shear momentum, putting together the data of mean velocity, Reynolds stress, and solid contribution. Indeed, the total shear stress in a channel with an elastic wall is the sum of three contributions: the mean viscous stress $\mu \mathrm{d}\bar{u}/\mathrm{d}y$, the off-diagonal component of the Reynolds stress tensor $-\rho \overline{u^\prime v^\prime}$, and the elastic shear stress $G\bar{B}_{xy}$ (which is null for the rigid wall case). 
Figure~\ref{fig: TotalRSS} shows the wall-normal profiles of all these components: in particular, the solid, dashed, dashed-dotted, and dotted lines represent the total, viscous, Reynolds, and elastic stress, respectively. The figure shows the data for the case with the elastic wall in the top row, and a rough wall in the bottom row. As expected from the values of the friction Reynolds number, when the wall deformation increases, the total stress near the walls increases for all cases, with the elastic one exhibiting a larger degree of enhancement. Consistent with the previous observations from the mean velocity profile (Figure \ref{fig: MeanVel}), the viscous shear stress is reduced with the presence of the elastic wall, a reduction compensated by an increase in the Reynolds and elastic shear stresses; a similar trend was also observed by \citet{Ardekani_Rosti_Brandt_2019} and by \citet{rosti2020low}. Different from the trend of the diagonal components of the Reynolds stress tensor (Figure \ref{fig: RSS}), the peaks of the viscous and Reynolds shear stress show no-clear shift towards the bulk region as the elasticity increases: the peak of the Reynolds shear stress is located at $(2-y)/h=\Tilde{y}/h \approx 0.2$ for both low and high cases; those of the viscous stress are located at $\Tilde{y}/h=0.07$ for low case, and $\Tilde{y}/h\approx 0.1$ for high case. However, we would like to point out that the peak location could be affected by the scaling of the wall-normal distance itself. \citet{rosti2020low} reported that the peak of the viscous stress shifts towards the fluid region, and that of the Reynolds shear stress shifts to the wall region; these profiles were plotted as a function of the wall-normal distance normalized by the wall-normal location of maximum streamwise velocity. \citet{Ardekani_Rosti_Brandt_2019} shows that the peak of the viscous stress is away from the wall, while the ones of the Reynolds shear stress are almost located at the same position; these profiles were illustrated in a symmetric computational domain. Our domain is asymmetric, and Figure \ref{fig: TotalRSS} shows that the position of zero total shear stress shifts to the bottom rigid wall side with increasing wall deformation. Similar to elastic walls, the viscous shear stress is reduced, and the Reynolds shear stress increases for the rough walls, however the changes are less than with the elastic wall; also, the peak of stress components is not significantly displaced when changing the level of roughness; the Reynolds shear stress peak is located at $\Tilde{y}/h=0.18$ among both roughness degrees; the ones of the viscous stress appears at $\Tilde{y}/h=0.016$ for the low case and  $\Tilde{y}/h=0.058$ for the high case.

In conclusion, the increased drag observed in both the elastic and rough wall cases can be attributed to a significant increase in the Reynolds stresses, although accompanied by a reduction of the viscous shear stress due to the weakening of the wall blocking and viscous effects. Notwithstanding these qualitative similarities, the two walls produce significant differences in the rest of the turbulent statistics. Indeed, the synergistic effect of wall undulation and elastic wall motion leads to stronger fluctuations, especially in the wall-normal direction, as well as an increase of the Reynolds and elastic stresses, thus enhancing the momentum exchange across the whole channel \citep{pope2001turbulent} and increasing the drag \citep{fukagata2002contribution}. The above results also show that the peak velocity perturbations move away from the wall towards the center of the channel, leading to a wide region of intense turbulent activity throughout the channel, see also \citet{Rosti_Brandt_2017} and \citet{rosti2020low}. These flow modifications cannot be explained uniquely in terms of wall undulations, as they are not present in the rough wall case, and suggest profound modifications of the turbulent structures and of the wall cycle, as will be discussed in the next section. 
 
\subsection{Turbulent structures} \label{sec: turbulent structure of e and r}
\subsubsection{Streamwise structures}
In this section, we focus on assessing the effect of the wall motion and undulation on the coherent turbulent structures, first by means of flow visualization and next through the quadrant analysis and autocorrelation functions.

\begin{figure}
  \centering
  \includegraphics[width=1\textwidth]{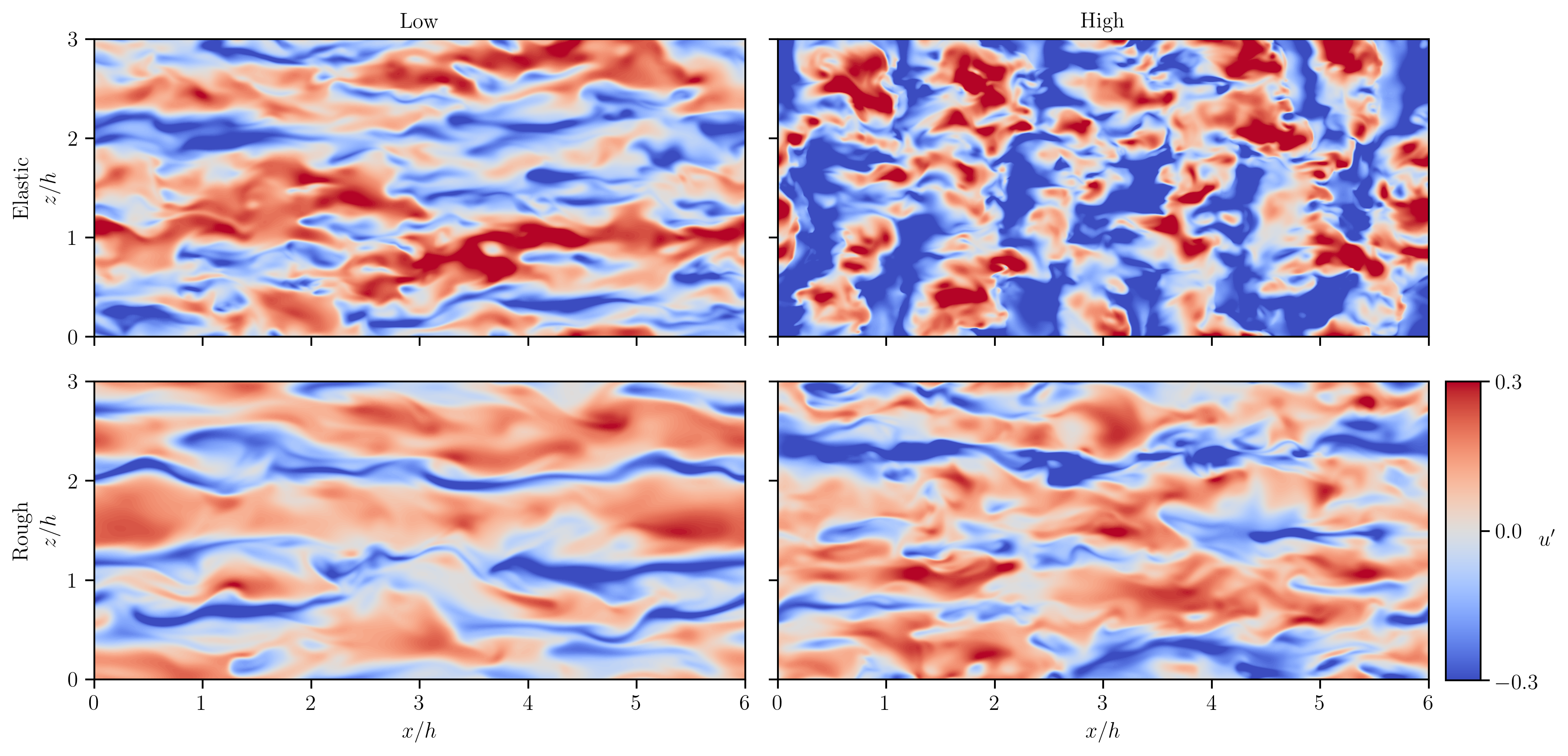}
  \caption{Instantaneous contour of the streamwise velocity fluctuations $u'$ in the $x-z$ plane at $(2h-y) = 0.2h$, in the elastic (top) and rough (bottom) cases with low (left) and high (right) wall deformation. The flow direction is from left to right.}
  \label{fig: streak}
\end{figure}

Figure~\ref{fig: streak} shows the streamwise velocity fluctuations in the $x-z$ plane at the wall-normal location $2h-y=0.2h$. This region is far from the average complex wall interface ($y=2h$), and it was chosen to ensure that the solid phase doesn't reach the plane. Although the wall shape and the bulk statistics, i.e., the drag increase due to the enhancement of the Reynolds stresses accompanied by a weakening of the viscous shear stress, between the elastic and rough cases are qualitatively similar, Figure~\ref{fig: streak} shows striking differences between the turbulent structures populating the region above the two walls. Indeed, as the wall elasticity increases in the first row of Figure~\ref{fig: streak}, the typical low- and high-speed structures become more fragmented and shortened, and large-scale spanwise coherent structures are formed. This modulation of streamwise structures is consistent with the previous observation made from Figure~\ref{fig: RSS}, where a relatively weaker growth rate in the streamwise direction was seen compared to the other two directions, thus implying a partial recovery of isotropy of the flow fluctuations.
In the rough case, instead, the spanwise elongated structures are not formed, and the only qualitative difference between the case with low and high deformation is the streak partial fragmentation and shortening. The behavior over rough walls is consistent with past results from the literature; for example, \citet{Narayanan_2024_direct} reported that turbulence over irregular rough surfaces exhibits more fragmented streaks compared to smooth surfaces. On the other hand, the flow structures observed for the elastic case seem to be a prerogative of the fully coupled case, in which the wall is allowed to move by the fluid-solid interaction. Shorter streaks and enhanced spanwise coherency of the turbulence structures were found by several authors when studying permeable walls and canopy flows, see e.g.,~\citet{suga2018anisotropic, kuwata2019extensive, gomez2019turbulent, kuwata2022role} and \citet{raupach1996coherent, nepf2012flow, sharma2020turbulent,monti2022solidity,  Foggi_Rota_Monti_Olivieri_Rosti_2024}, with a possible explanation coming from the presence of significant wall-normal fluctuations at the wall/interface.

\begin{figure}
  \centering
  \includegraphics[width=0.8\textwidth]{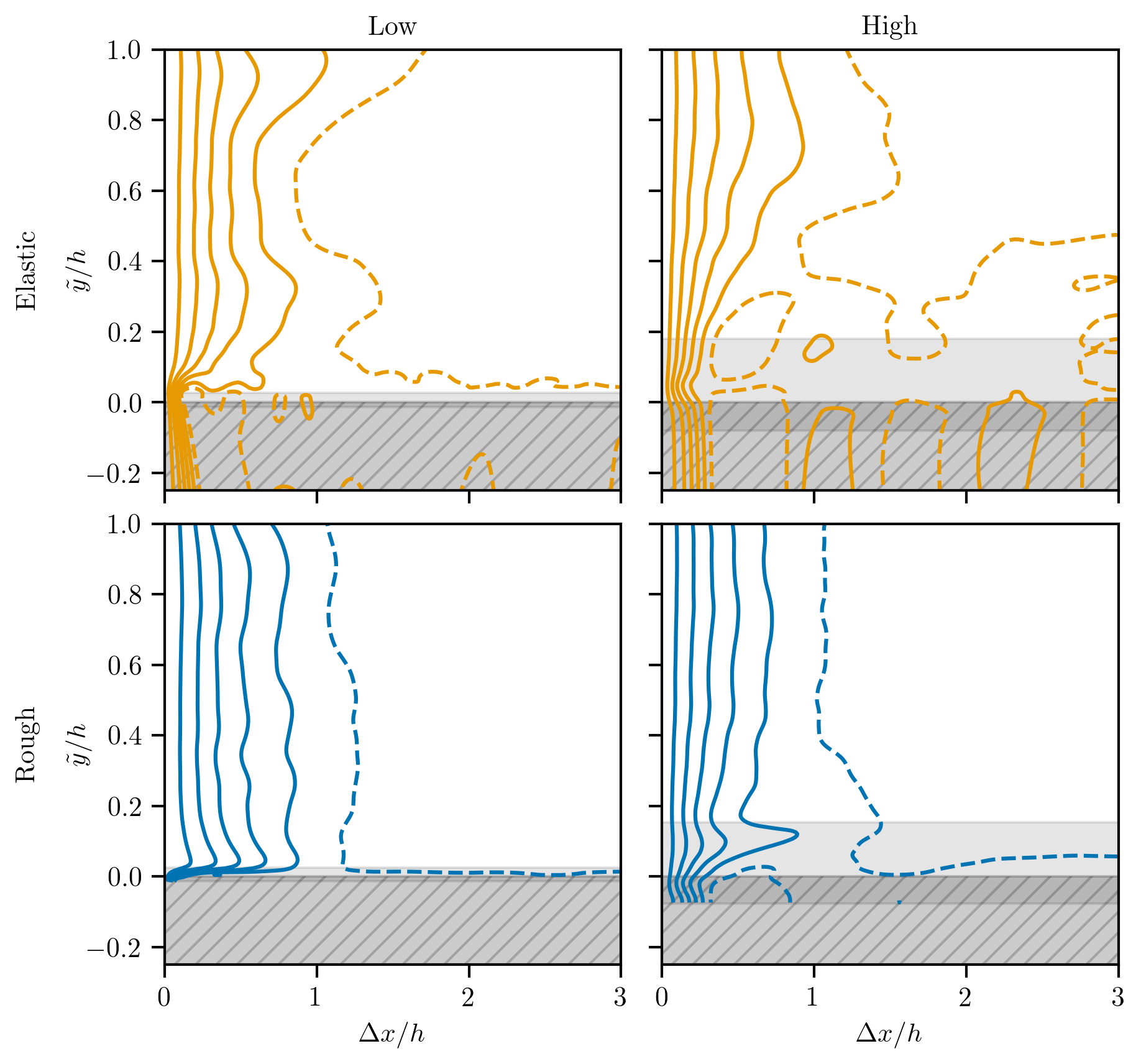}
  \caption{One-dimensional autocorrelation functions of the streamwise velocity fluctuations as a function of the streamwise separation, stacked for different wall-normal distances from the top wall, i.e., $\Tilde{y}=2h-y$. The top and bottom rows show the results for the elastic and rough cases, and the columns represent the level of surface deformation: low (left) and high (right). The shaded gray area with hatching shows the region occupied by the solid, while the region without hatching represents the area spanned by the wall fluctuations. The color lines represent the values from $-0.1$ to $0.9$, with $0.2$ increments, and the dashed and solid lines are used to distinguish the negative and positive values.}
  \label{fig: space Ruux}
\end{figure}

\begin{figure}
  \centering
  \includegraphics[width=0.8\textwidth]{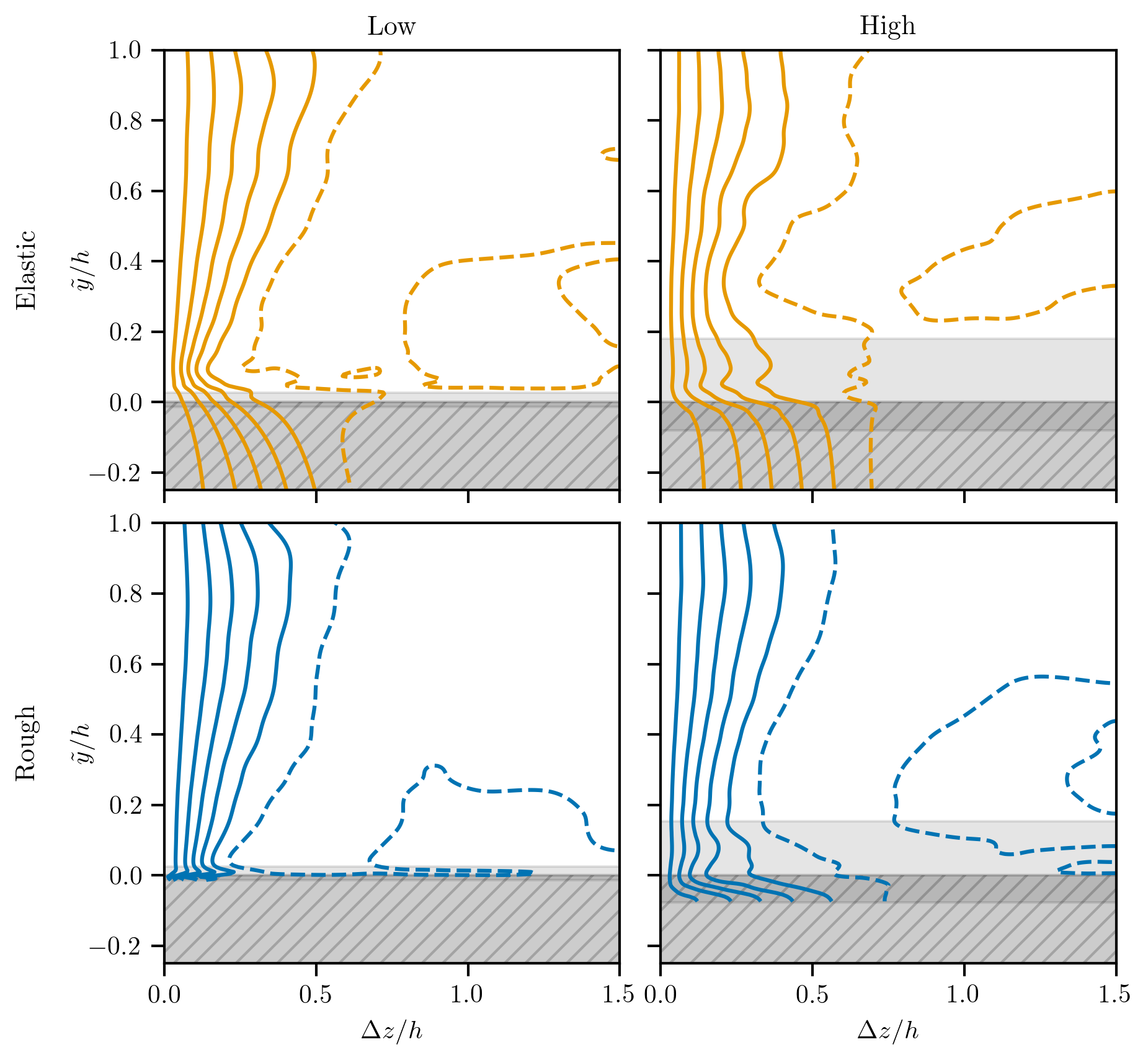}
  \caption{One-dimensional autocorrelation functions of the streamwise velocity fluctuations in the spanwise direction. The details of the figure are the same as in the previous figure.}
  \label{fig: space Ruuz}
\end{figure}

\subsubsection{One-dimensional autocorrelation functions}
Next, we quantify these observations by means of the one-dimensional autocorrelation function of the velocity fluctuations, defined as
\begin{equation}
    R_{u_i^\prime u_i^\prime,x_k} = \frac{\overline{u'_i(x_k)u'_i(x_k+\Delta x_k)}}{\overline{u'_i(x_k)u'_i(x_k)}}, 
    \label{eq: correlation func}
\end{equation}
where the bar represents average, and $\Delta x_k$ denote the spatial separation. Figure~\ref{fig: space Ruux} shows the one-dimensional streamwise autocorrelation of the streamwise velocity fluctuations $u^\prime$. The effect of increasing wall deformation is clearly visible in the figure, as well as the big differences between the rough and elastic cases. For example, if we focus on $\Tilde{y}=0.2h$ (the same wall-normal position as Figure \ref{fig: streak}), we can observe that the correlation is comparable in the two configurations when the deformation is small. However, for larger wall undulations, it decreases for both cases; this effect is more prominent for the elastic wall, consistent with Figure~\ref{fig: streak}. Additionally, local positive and negative values are observed along the streamwise direction for both cases, in the region partially occupied by the solid when the wall undulation is large enough. Such cell-like distinctive patterns were also observed by \citet{Breugem2006The} when studying turbulent flows over a porous medium, and were attributed to large-scale pressure fluctuations. In our case, the characteristic wavelength of this pattern is approximately $h$, which corresponds to the wavelength of wall undulation found in \S~\ref{subsec: wall configuration}. Although the surface undulations are comparable among the elastic and rough cases, this distinctive pattern is most prominent in the elastic case, thus being promoted by the wall motion on top of the wall shape effect. To complete the picture, Figure~\ref{fig: space Ruuz} presents the spanwise autocorrelation of the streamwise velocity fluctuations. Focusing again at $\Tilde{y}=0.2h$, we can observe that the high elastic case shows a significantly increased spanwise correlation (consistent with Figure~\ref{fig: streak}), while the effect of roughness is minor, although still present. Note that the level of correlations remains large in the whole region swept by the wall motion (the gray region). On the other hand, the rough case shows a correlation that rapidly decreases in the same region (when moving away from the wall), re-emphasizing the fact that it is the dynamic wall motion that enhances the spanwise coherency.

\begin{figure}
    \centering
    \includegraphics[width=0.7\textwidth]{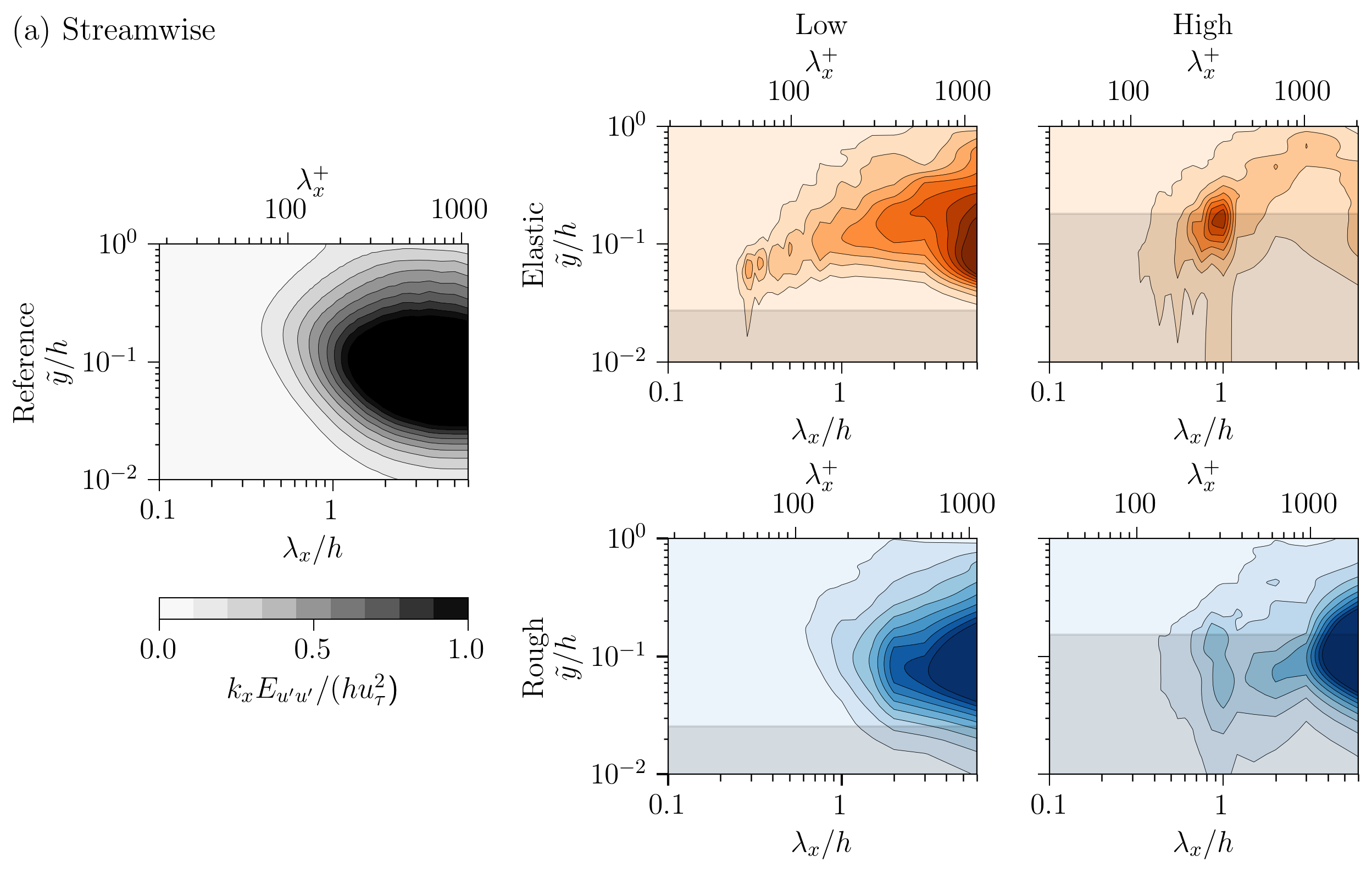}
    \includegraphics[width=0.7\textwidth]{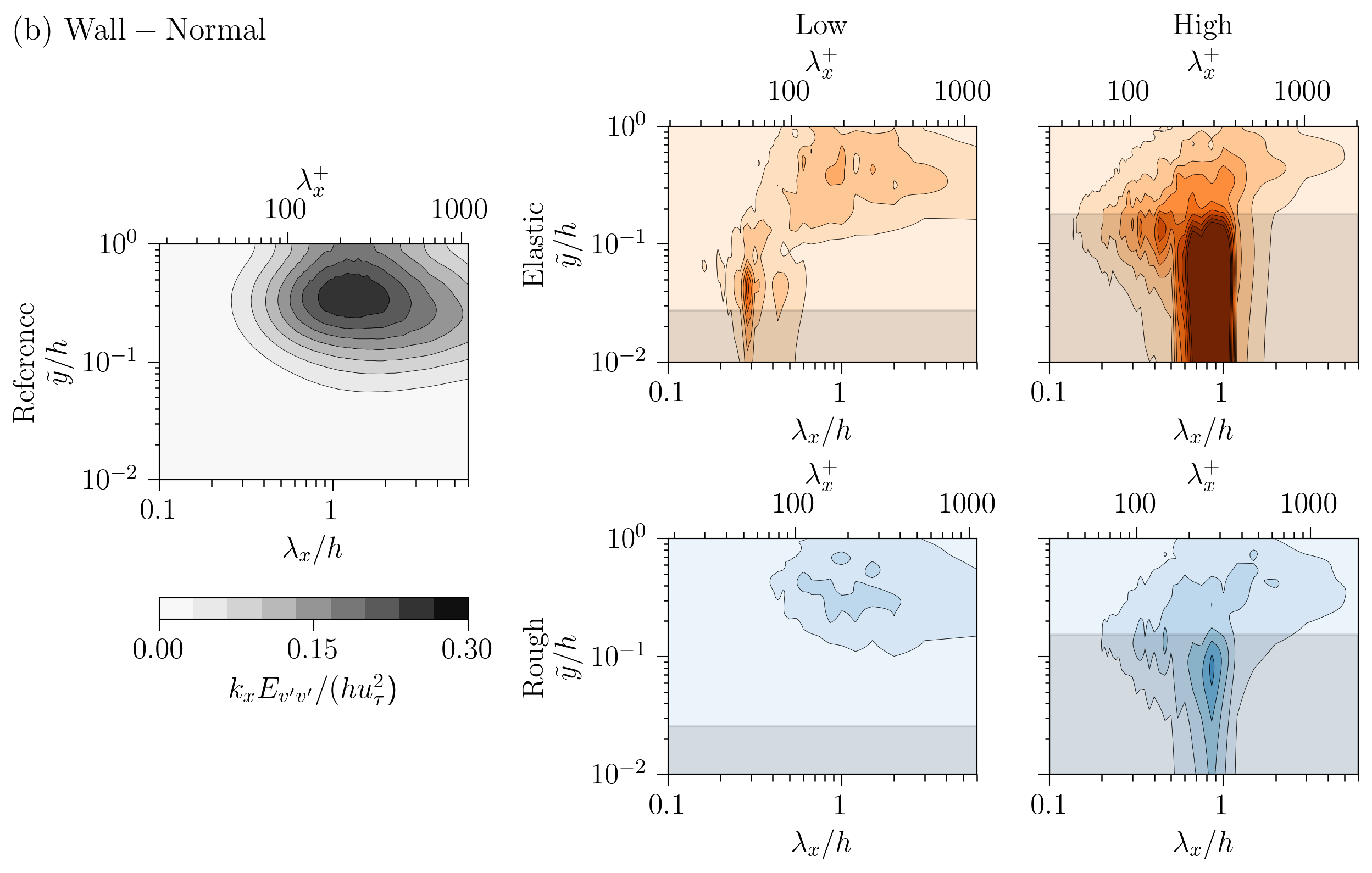}
    \includegraphics[width=0.7\textwidth]{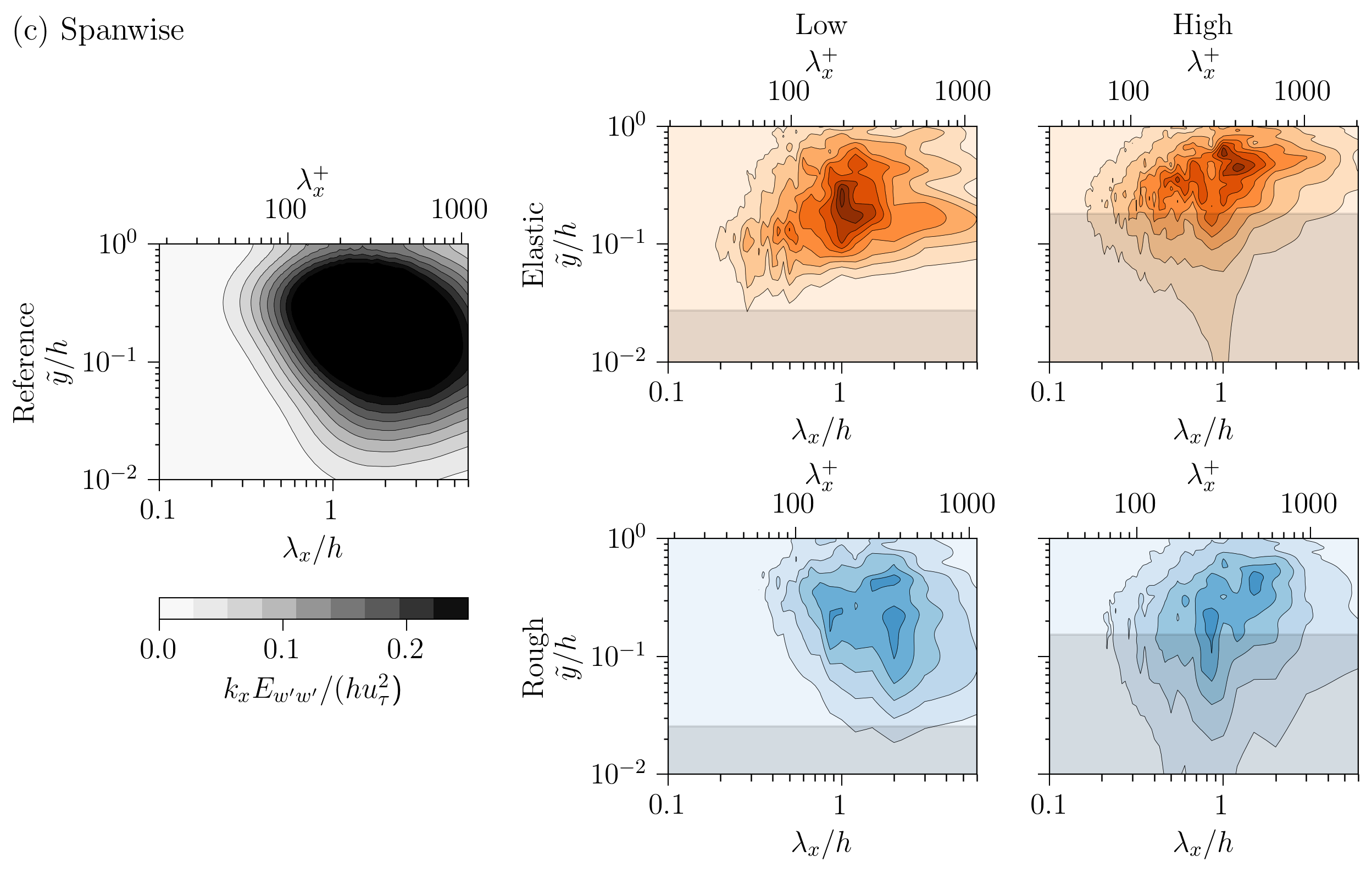}
    \caption{The premultiplied spectra, $k_x u_i'^2/(hu_\tau^2)$ in the streamwise-wall-normal plane. The leftmost black plot refers to the smooth channel case, while the orange and blue ones correspond to the elastic and rough wall cases, respectively, shown for both the low (left) and high (right) degrees of deformation. The contour level range is the same for all cases for each velocity component. The shaded gray region represents the area spanned by the wall undulations. The spectrum data for the plane channel case is taken from \cite{lee2015direct}.}
    \label{fig: energy spectra st}
\end{figure}

\begin{figure}
    \centering
    \includegraphics[width=0.7\textwidth]{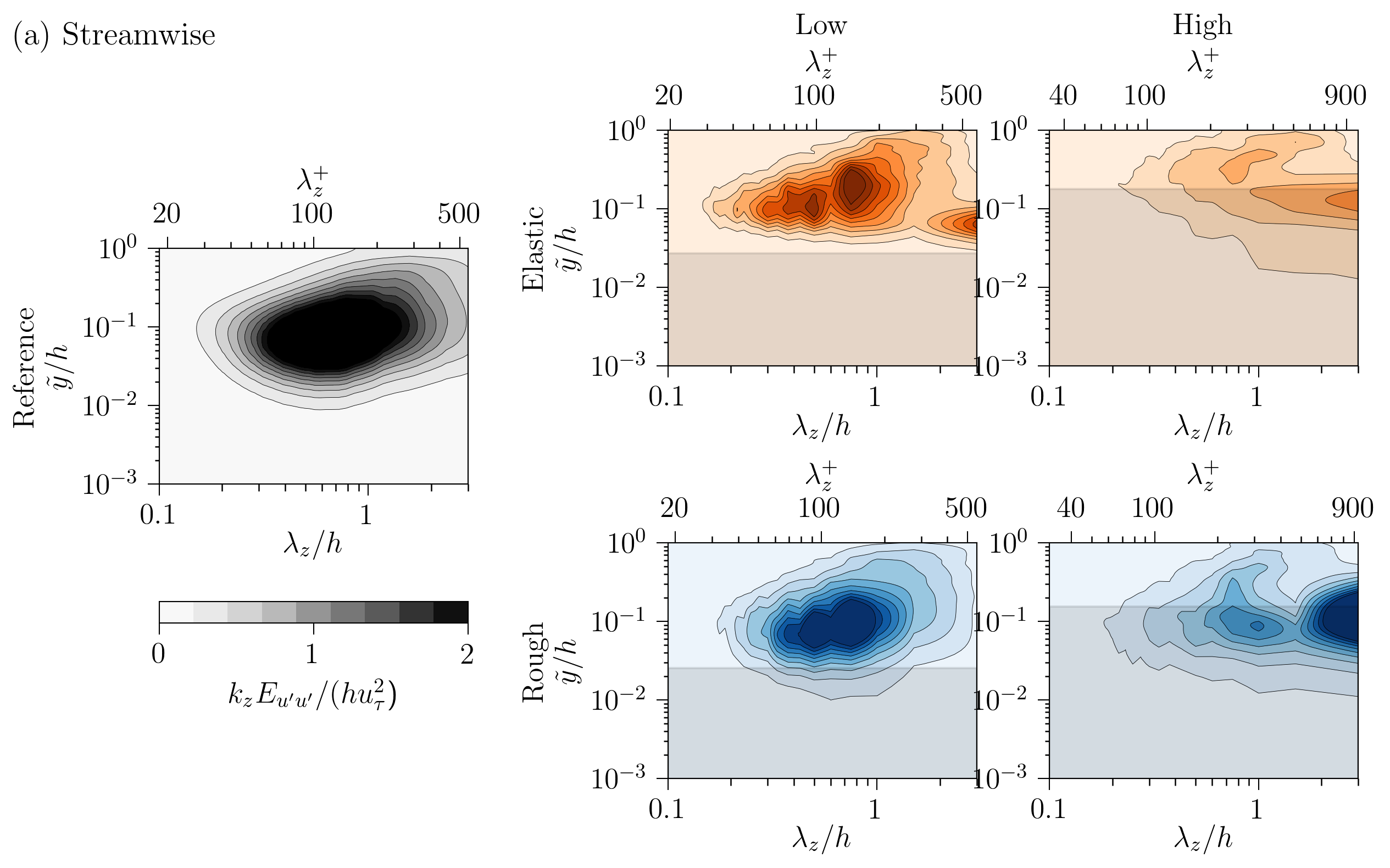}
    \includegraphics[width=0.7\textwidth]{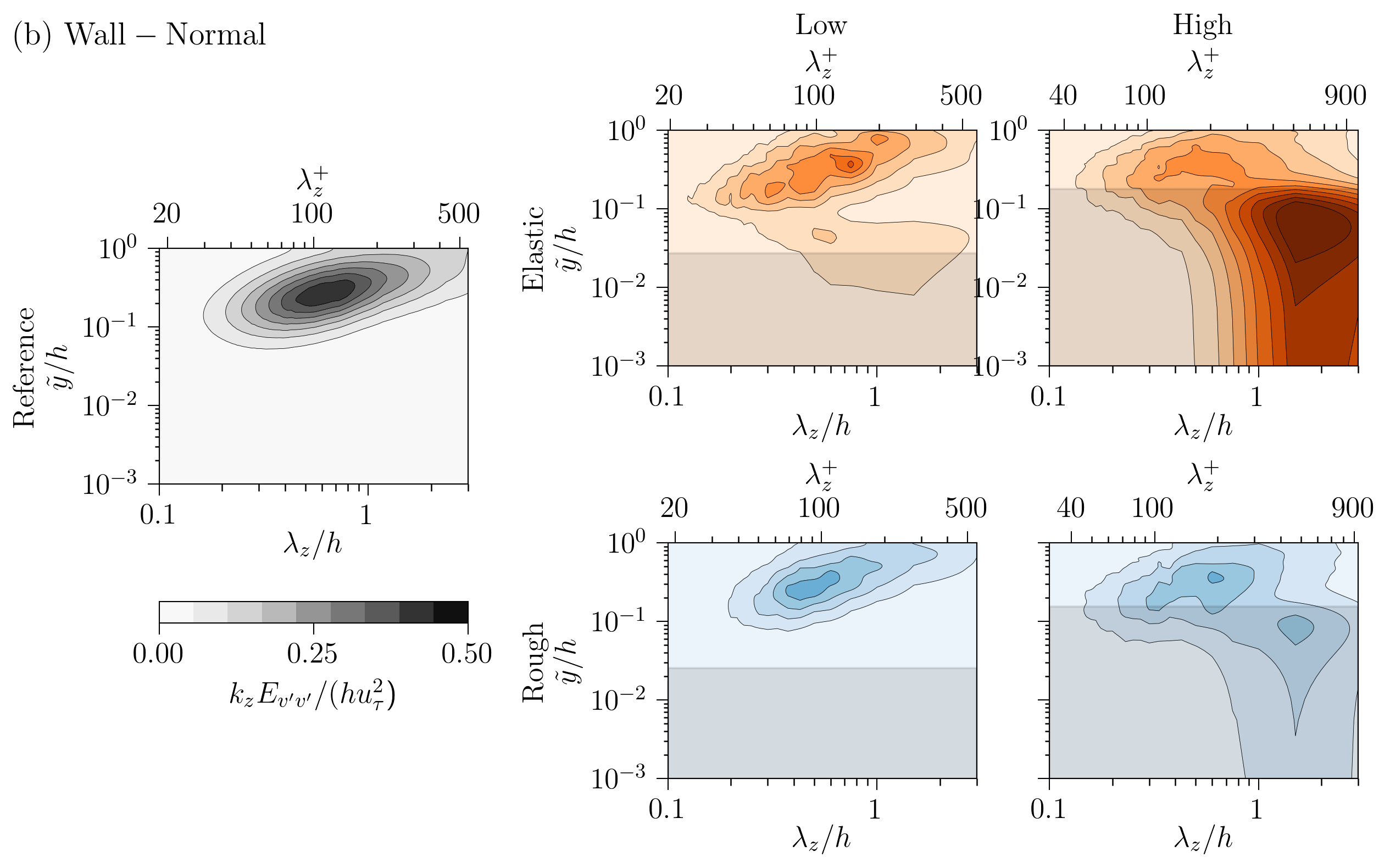}
    \includegraphics[width=0.7\textwidth]{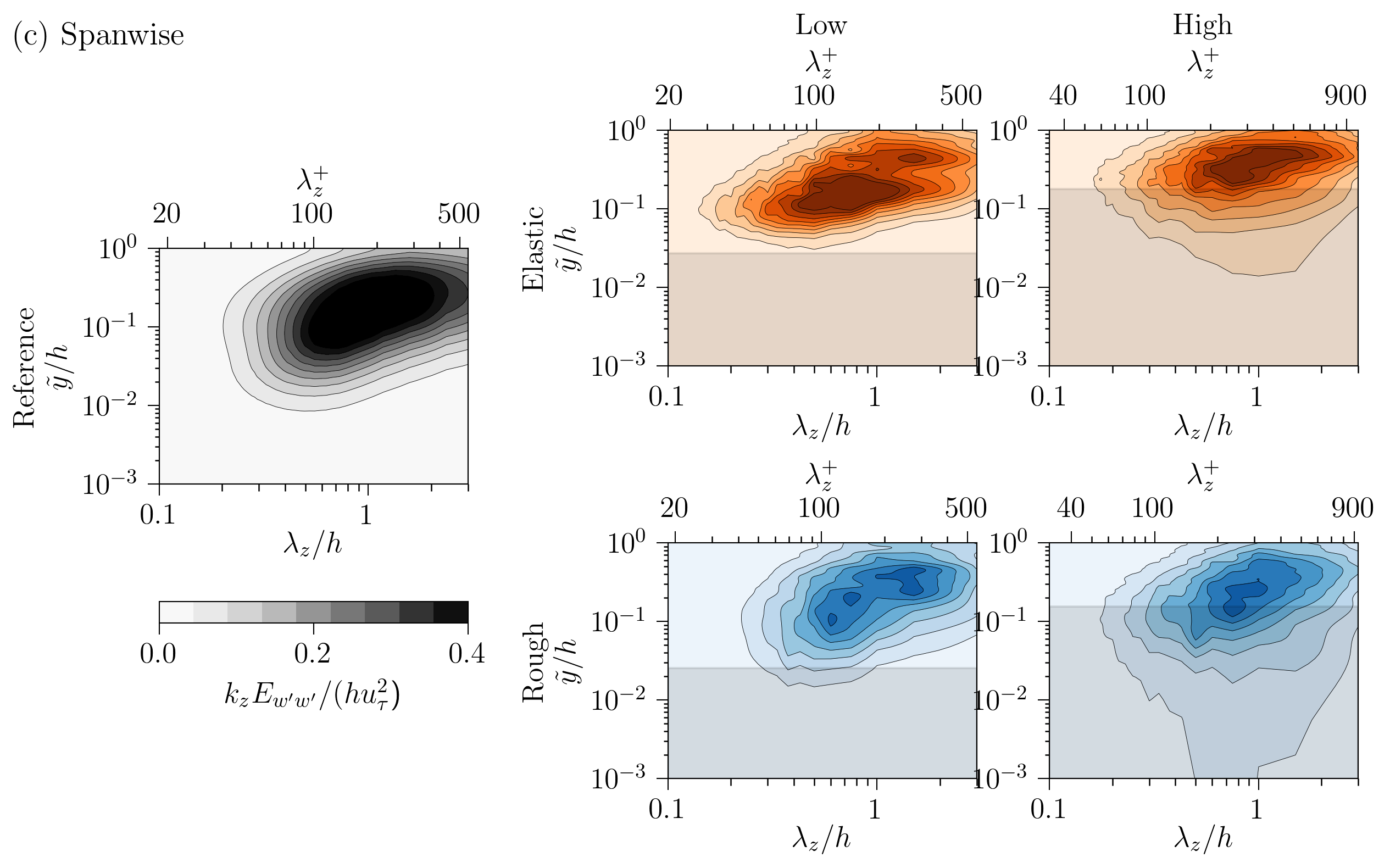}
    \caption{The premultiplied spectra, $k_z u_i'^2/(hu_\tau^2)$ in the spanwise direction. The leftmost black plot refers to the smooth channel case, while the orange and blue ones refer to the elastic and rough wall cases, respectively, shown for both the low (left) and high (right) degrees of deformation. The contour level range is the same for all cases for each velocity component. The shaded gray region represents the area spanned by the wall undulations. The spectrum data for the plane channel case was taken from \cite{lee2015direct}.}
    \label{fig: energy spectra sp} 
\end{figure}  

\subsubsection{Premultiplied energy spectrum of the velocity fluctuations}
To characterize the predominant length scale observed in the coherent structures and the autocorrelation functions, we look at the premultiplied energy spectrum of the velocity fluctuations, $k_{i} u_k'^2/(hu_\tau^2)$, where $k_{i}$ is the wavenumber with $i$ being the $x$ or $z$ directions. Figures~\ref{fig: energy spectra st} and Figure~\ref{fig: energy spectra sp} show the iso-contour of the premultiplied energy spectra as a function of the wavelength $\lambda_i/h$ and wall-normal distance $\tilde{y}=2h-y$, with $i$ being in streamwise (a), wall-normal (b) or spanwise (c) directions. The colors of contours refer to the smooth (black), elastic (orange), and rough (blue) wall channels, respectively.

We start by looking at the streamwise spectra, reported in Figure~\ref{fig: energy spectra st}. The smooth wall case (black-colored contour) exhibits a clear single peak in the distributions, which is comparable to what was observed for the high-roughness case, but with reduced intensity. Figure~\ref{fig: energy spectra st} shows that in the elastic case, the dominant length scale becomes smaller in the streamwise direction with an increase in the wall elasticity. Indeed, the elastic wall cases exhibit a peak closer to the wall and at a smaller length scale, at around $\lambda_x/h \approx 1.0$ for all velocity components. This is also the length scale that is observed in the instantaneous contour plot of the streamwise velocity fluctuation (shown in Figure~\ref{fig: streak}) and the wavelength of the surface undulations (shown in Figure~\ref{fig: psd}). 

Next, we consider the spanwise premultipled energy spectra, as shown in Figure~\ref{fig: energy spectra sp}. For the rigid wall case, the streamwise fluctuations are dominant compared to other velocity components, and the peak of the wall-normal components appears slightly farther away from the wall rather than the stream and span components. With the increase in wall elasticity, Figure~\ref{fig: energy spectra sp} shows that the peaks in the streamwise and wall-normal directions are distributed around larger length scales, typical of large-scale rollers. In both the streamwise and spanwise directions, the wall-normal component exhibits intense values also within the area spanned by the wall fluctuations (gray colored region in the figure). Similar distribution is also observed in the low-roughness case, but overall, the spectrum distributions correspond to those of the smooth wall case except for the streamwise components. As shown in Figure~\ref{fig: energy spectra sp} (a), there is an enhanced correlation along the span, which is representative of the turbulent structures getting slightly bigger along that direction.

From the above observations, the motion of the elastic walls is seen to influence the streamwise length scale by weakening it. Eventually, the spanwise length scale gets larger, and the velocity fluctuations (especially the wall-normal component) in the range of wall-motion get amplified by the elastic wall.  This observation is also reported by \cite{luhar2015framework}, who said that the large, steady spanwise mode can be strengthened efficiently over the compliant wall.

\subsubsection{Quadrant analysis}
As a final step, we use the quadrant analysis to identify the impact of the complex walls on the different events \citep{Wallace_Eckelmann_Brodkey_1972,wallace2016quadrant}. Each quadrant is defined as
\begin{equation}
    Q_m = \frac{1}{N_m}\sum(u^\prime \Tilde{v}^\prime)_m,
 \label{eq: quadrant}
\end{equation}
where $N_m$ is the number of events in each quadrant $m=1,2,3,4$. The first $Q1$ ($u^\prime > 0$, $\Tilde{v}^\prime > 0$) and the third $Q3$ ($u^\prime < 0$, $\Tilde{v}^\prime < 0$) quadrants contribute to the positive production of the shear Reynolds stress, and have a minimum contribution to the Reynolds shear stresses. The second $Q2$ ($u^\prime < 0$, $\Tilde{v}^\prime > 0$) and the fourth $Q4$ ($u^\prime > 0$, $\Tilde{v}^\prime < 0$) quadrants represent ejection and sweep events, and are significant contributors to the turbulent kinetic energy. 
Note that, to compute $Q_m$ we use only the fluid velocity fluctuations, and that we consider the wall-normal velocity $\Tilde{v}'$, corresponding to the $\Tilde{y}=2h-y$ coordinate system, where a positive $\Tilde{v}'$ corresponds to a fluctuation directed from the top-complex wall towards the inner fluid region.

\begin{figure}
  \centering
   \includegraphics[width=0.9\textwidth]{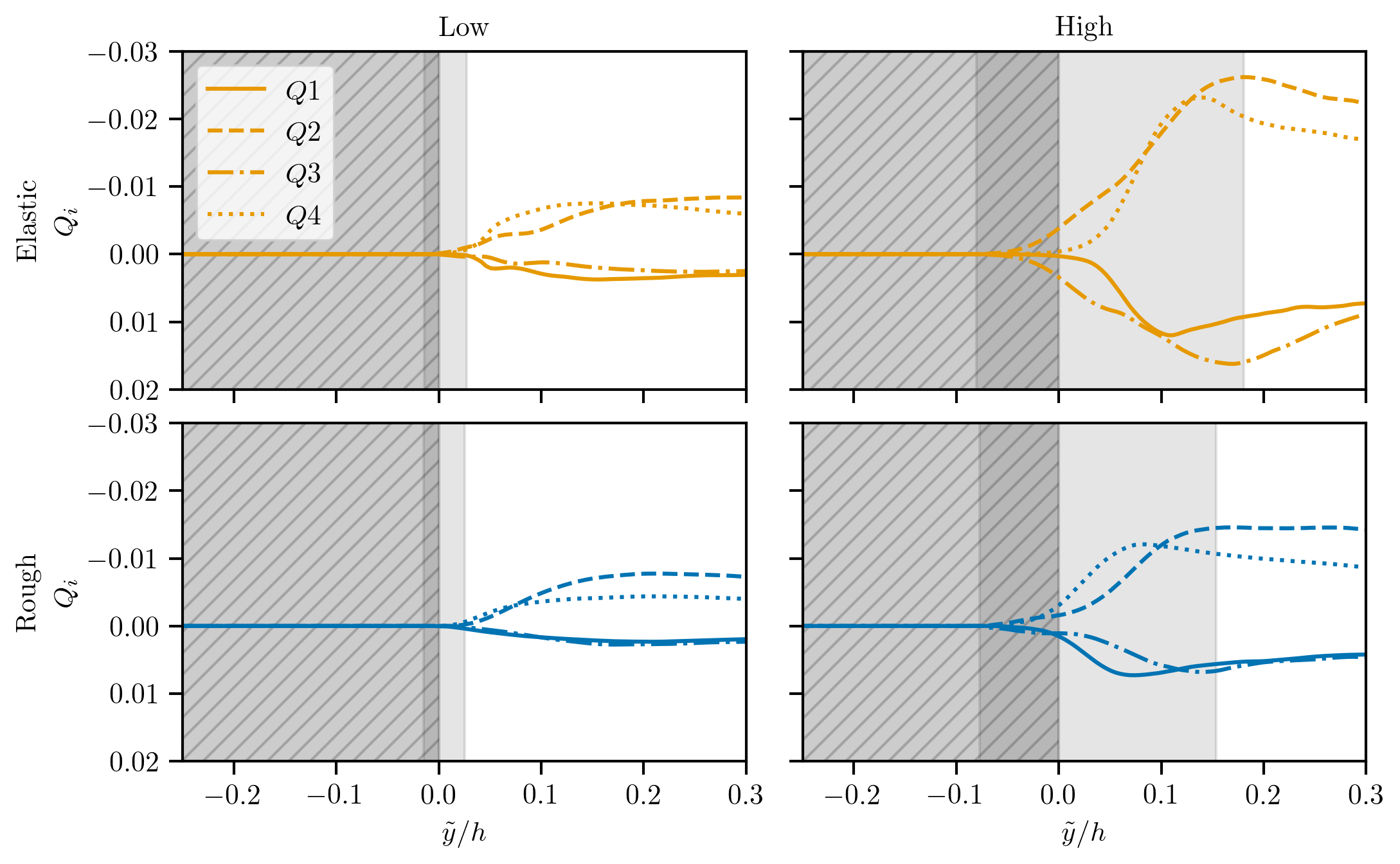}
  \caption{Frequency of the events of the four quadrants of the Reynolds shear stress, as a function of the wall-normal direction. The top panels correspond to the elastic wall, and the bottom to the rough one, while the two columns show increasing levels of wall deformation, going from left to right. The line styles distinguish the quadrant: $Q1$ (continuous), $Q2$ (dashed), $Q3$ (dash-dotted), and $Q4$ (dotted). The background gray with hatching shows the average solid layer, with the minimum and maximum extension of the solid marked in plain gray.}
  \label{fig: quadrant} 
\end{figure}

Generally, planar turbulent channel flows show predominant sweeps close to the wall, ejection away from the wall, and with ejection and sweep contributions being approximately the same at around $12$ wall units \citep{kim1987turbulence}. Figure~\ref{fig: quadrant} shows the decomposed quadrant events as a function of the wall-normal direction $\Tilde{y}$, for the low and high complexity cases of the elastic and rough walls.
While the low elasticity case exhibits a trend similar to the classical turbulent channel flows, the situation changes when elasticity increases, with all events becoming generally stronger and propagating deeply inside the solid layer and in the bulk of the channel, consistent with the trend of velocity fluctuations shown in Figure~\ref{fig: RSS}.
The high-elasticity case shows that $Q2$ (ejections) events significantly contribute to turbulent activity within the region of oscillation of the elastic wall, rather than the usual $Q4$ (sweeps) events, and we also observe quite strong $Q3$ events. $Q2$ is the predominant event not only within the elastic wall oscillation region, but also in the bulk, where $Q2$ events overtake $Q4$ ones also in classical channel flows.

For rough walls, instead, the trend of the quadrants is similar to that of planar channel flows, as shown in the bottom rows of Figure~\ref{fig: quadrant}. Indeed, $Q4$ dominates over $Q2$ near the wall, with the strength of each event increasing with the amplitude of the roughness. The wall-normal position where $Q2$ motions overtake $Q4$ moves away from the interface for large roughness, as the wall undulations increase. As expected, inside the completely rigid layer, both motions vanish entirely. The above trends are also observed in the study by \citet{COCEAL_DOBRE_THOMAS_BELCHER_2007}, where turbulent flows over a three-dimensional array of cubical roughness were investigated; the authors found that strong sweeps were predominant in the rough region, while strong ejections were dominant above the top of the array, consistent with our results. Also, it was found that two-dimensional roughness leads to the reduction of ejections \citep{krogstadt1999surface}, caused by the low-momentum fluid being trapped between the rough elements \citep{krogstad2005experimental}. Similar trends have been observed also for other complex walls, such as porous media \citep{Suga2011vortex,kuwata2016transport,li2019drag} and canopy flows \citep{finnigan2000turbulence, poggi2004effect}, where it was reported that the turbulent activity is dominated by strong $Q4$-sweeps close to the interface, and dominating $Q2$-ejections far away from the interface. In addition, \citet{kuwata2016transport} compared the turbulent flow over a rough wall with a rough-permeable wall, with identical interface geometry. Both cases showed more decisive sweep than ejection events at the interface, especially for the rough-permeable walls, with the location where the ejection events overtake being roughly the same at around $13$ wall units, as also shown by \citet{Suga2011vortex, kuwata_suga_2017_direct, finnigan2009turbulence}.

Overall, the rough walls show a behaviour that does not significantly change from that of the classical channel flows, unlike what we observed for the channels with elastic walls, where the motion of the elastic walls strongly modifies the distributions of the flow fluctuations close to the wall and throughout the channel.

\begin{figure}
    \centering
    \includegraphics[width=0.9\textwidth]{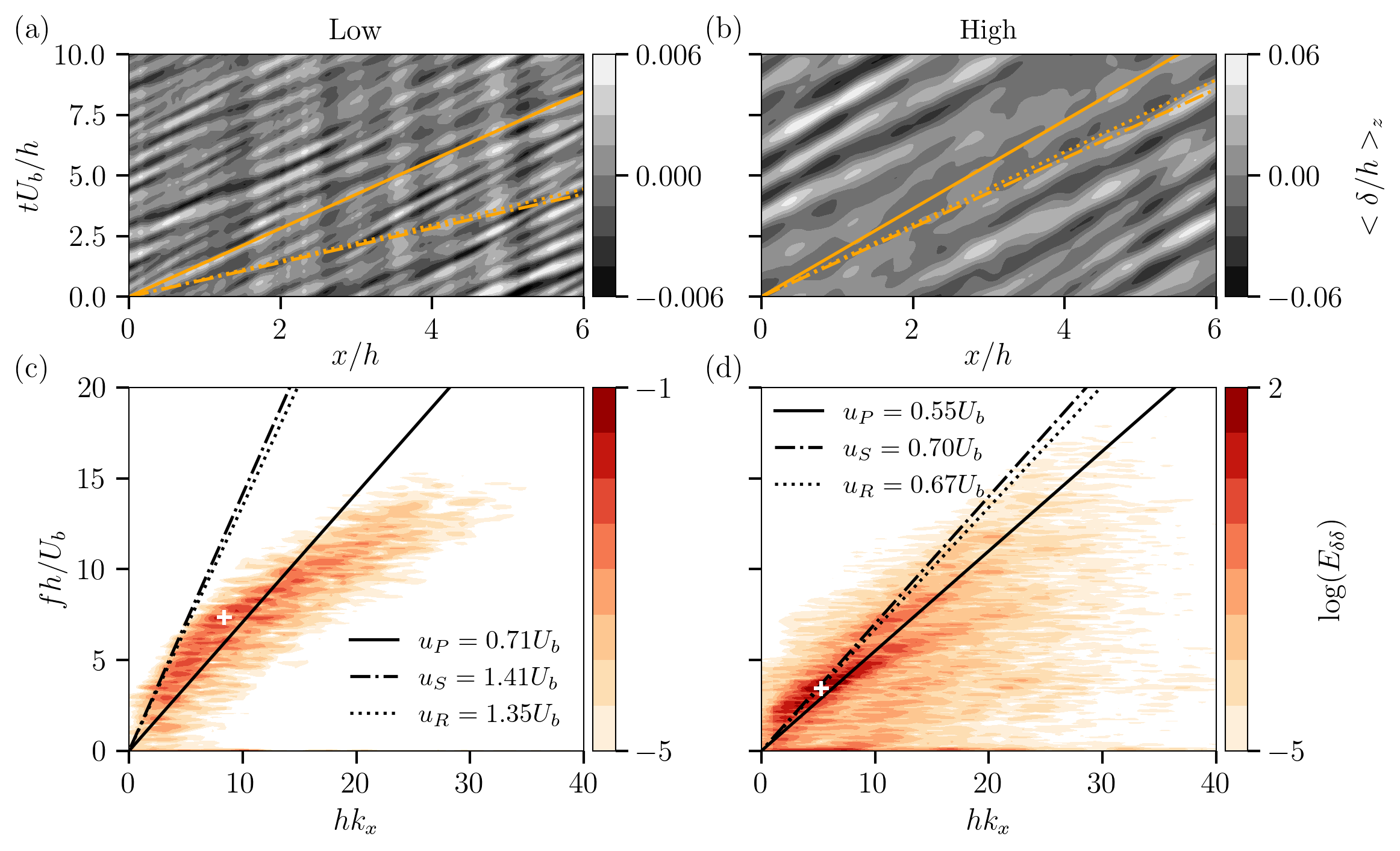}
    \caption{Physical- and Fourier-space time diagram of the amplitude of the elastic wall interface, averaged along the spanwise direction, for the (left, -a, -c) low- and (right, -b, -d) high-elasticity cases. (a) and (b) present the physical space-time diagram. (c) and (d) present streamwise wavenumber-frequency spectra. The lines represent different shear speeds: (solid) calculated from the present data, (dashed-dotted) the Rayleigh wave $u_R = 0.954\sqrt{G/\rho_s}$, and (dotted) the analytical shear speed $u_S = \sqrt{G/\rho_s}$. The white cross represents the position corresponding to the peak of the spectrum.}
    \label{fig: TX_FKXspectrum}   
\end{figure}

\subsection{Spatio-temporal analysis of the elastic wall displacements}
To understand the links between the motion of the elastic wall on the spatial and temporal length scales of the flow, we examine the space and time variation of the elastic wall displacements for the low and high elasticity cases. Figures~\ref{fig: TX_FKXspectrum} (a) and (b) show the displacement of the elastic wall as a function of space along the streamwise direction (averaged over the spanwise direction) and time. The surface displacement propagates in the streamwise direction with a specific speed $u_{P}$, marked by the solid line. In the figure, the dashed-dotted line represents the Rayleigh wave speed $u_R = 0.954\sqrt{G/\rho_s}$ and the dotted line represents the speed of the shear waves $u_S= \sqrt{G/\rho_s}$. Additionally, we show the streamwise wavenumber spectrum in Figure~\ref{fig: TX_FKXspectrum} (c) and (d). These spectra are calculated by calculating the Fourier transform of the temporal data in the streamwise direction, at each spanwise location and then subsequently averaging across the spanwise direction. As elasticity increases, the calculated shear speed $u_P$ reduces from $0.71U_b$ to $0.55U_b$. This trend was reported by \cite{Esteghamatian2022spatiotemporal} (numerical) and \cite{lu2024scaling} (experimental), the opposite trend by \cite{greidanus2022response}. When spanwise structures are dominant, the propagation speed is $0.53U_b$ in \cite{lu2024scaling}, $0.55U_b$ in the present study, and $0.65U_b$ in \cite{Esteghamatian2022spatiotemporal}. The speed of the streamwise traveling wave $u_P$ becomes closer to the material shear speed $u_S$ as the elasticity increases, and thus Figure~\ref{fig: TX_FKXspectrum} (d) shows a better correlation than (c). This is consistent with the previous observations that only the case with high elasticity exhibits the strongly spanwise coherent structures. Consistently with \cite{Esteghamatian2022spatiotemporal}, the observed speed is always slower than the pure shear speed $u_S$, and more similar to the Rayleigh wave advection speed $u_R$, even with different elasticity. Also, we computed the same in the spanwise direction, but no clear propagation mode is evident (not shown). Overall, these observations agree with the results by \cite{Esteghamatian2022spatiotemporal}.

\section{Conclusions}\label{sec: conclusion}
This study aimed to understand the effects of wall undulations, wall-normal velocity fluctuations, and the dynamic wall movement on the flow physics of elastic walls. This was done by comparing the elastic walls (which is a combined scenario of all the above) with equivalent rough walls, and thus isolating these effects individually. We performed direct numerical simulations of turbulent channel flows over different types of complex walls and compared the results. The results show that the dynamic behavior due to the elastic walls contributes the most in determining its near-wall flow behavior, followed by wall undulations due to the roughness.

First, we consider elastic and rough walls to understand the effect of the dynamic fluid-solid interaction on the flow. Thus, we consider elastic walls of different flexibilities and equivalent rough walls, obtained from an instantaneous surface undulation of the elastic walls.
The bulk turbulent statistics show similar behaviours in a qualitative sense; both elasticity and roughness intensify turbulent activities. However, there are specific differences in the way these effects are manifested; the elastic wall changes the slope of the mean velocity profile considerably, and the Reynolds stresses peak away from the wall (more towards the bulk) compared to the rough wall. There is a partial recovery of isotropy of the velocity fluctuations observed for the elastic wall, where the wall-normal and spanwise components grow stronger than the streamwise one, with the magnitude of all components becoming almost comparable when the elasticity increases. Differently, for the rough wall, only the near-wall region is affected, and the observed trends are more similar to those of the classical channel flows, with the streamwise component still remaining the largest. The above observations suggest that, while turbulent flows over relatively rigid elastic walls show similar effects produced by the rough walls, when wall compliance increases, additional effects arise, which are peculiar to the fluid-structure interaction.
These differences can be explained in terms of turbulent structures, which exhibit clear differences for the rough and elastic walls. Indeed, as a result of the dynamic interaction between the fluid and elastic wall, there is an increased vertical momentum exchange in the fluid, resulting in the highly elastic walls showing a spanwise organized behavior, while the rough walls maintain in relative scales a predominant streamwise coherency. For rigid and smooth walls, sweeps dominate the near-wall region, with ejections taking over in the bulk of the channel. With increasing wall deformation, this remains unchanged for the rough wall, while the elastic-near-wall region is dominated by ejections, an effect induced by the wall motion. The intense wall-normal fluctuations, characterized by the strong ejections, and the cell-like patterns of the autocorrelation function in the streamwise direction $R_{u'u',x}$ observed for the elastic walls, are consistent with previous numerical and experimental results \citep{Rosti_Brandt_2017, Ardekani_Rosti_Brandt_2019, Esteghamatian2022spatiotemporal}. We also examine the speed at which waves propagate on the surface of the elastic wall. Even for low-elasticity cases, the wave propagates dominantly in the streamwise direction, with the shear speed decreasing and getting closer to the material shear speed with an increase in elasticity. These waves, clearly absent in the rough cases, can be a source of the observed differences among the elastic and rough walls.

In addition to the wall shape (evaluated through rough walls), we also conducted additional simulations to understand the effect of the strong wall-normal fluctuations at the interface (reported in Appendix~\ref{asec: porous} of the manuscript). For this, we compare turbulent flows over elastic and porous walls, the latter created with the same average vertical fluctuations at the interface. Compared to elastic walls, the mean-velocity profile of the porous walls is more similar to that found for classical channel flows; similarly, a comparable little effect on the drag is measured for the porous wall. In the latter, with observe that the turbulent structures are slightly strengthened along the streamwise and spanwise directions, in contrast to elastic walls producing a strong spanwise coherency. In other words, structural modifications similar to those obtained in the elastic cases cannot be observed by just introducing the mean wall-normal disturbances at the interface.

In conclusion, roughness accounts for a large part of the flow modifications observed for the elastic walls, but the coupled wall movement definitely introduces additional peculiar effects that further significantly alter the resulting flow. Although the wall-normal fluctuations are also enhanced by increasing the wall elasticity, the effect of the vertical fluctuation at the interface in itself cannot reproduce the structural changes observed in the elastic cases, re-emphasizing the need to include the full dynamic fluid-structure interaction effects when modeling the turbulent flow over an elastic wall. While it would be interesting to investigate even larger Reynolds numbers in the future, the computational cost can soon become prohibitive. Implementing simple models for the walls, e.g.,  a spring-mass-damper model~\citep{kim2014space}, a model formed by a tensioned network of compressive members interconnected by tensile members \citep{luo2005accurate} \textit{etc.} might help to reduce the computational cost, but would be subject to the limitations of the chosen model.

\appendix
\section{Continuum mechanics for the non-linear hyperelastic material}\label{asec: basics for continuum mechanics}
When continuum structures deform, the body changes in its shape and/or volume from an initial (undeformed) configuration to a current (deformed) configuration. To associate between both the coordinates, the deformation gradient tensor $\textbf{F}(\textbf{X},t)$ ($\textbf{F}$ hereafter) is introduced, defined as,
\begin{equation}
  \textbf{F} = \frac{d\textbf{x}}{d\textbf{X}},
\end{equation}
where, $\textbf{X}$ is a position vector in the undeformed configuration and $\textbf{x}$ is a position vector in the new deformed configuration. As a deformation tensor, the Green strain tensor (or left Cauchy-Green deformation tensor), $\textbf{B}$, is introduced as:
\begin{equation}
  \textbf{B} = \textbf{F}\textbf{F}^T.
\end{equation}

Further, a strain energy density function $W$ is constructed, which is a scalar-valued function that governs the relationship between stress and strain. Once $W$ is known, we can obtain the stress under any deformation. $W$ is a function of the invariants of $\textbf{B}$ and thus has the benefit that the obtained values are independent of the coordinate. Using $W$, the elastic part of the Cauchy stress is expressed by 
\begin{equation}
  \boldsymbol{\sigma} = \frac{1}{\det \textbf{F}}\frac{\partial W}{\partial \textbf{F}}\cdot \textbf{F}^T
\end{equation}
The details about the Green strain tensor can be found in e.g., \citet{bonet1997nonlinear} (chapter 3).

In the current study, we have modeled the solid as a hyperelastic neo-Hookean model. A hyperelastic material is a class of elastic materials for which the work done by the stresses is only dependent on the initial and final configurations, and the relation between stress and strain is described through a strain energy density function, $W$. A Mooney-Rivlin material is widely used for modeling rubber-like materials, whose stress-strain relationship is not linear, while the behavior is assumed to be completely elastic, isotropic, and incompressible throughout the deformation process. Particularly, a neo-Hookean model is a particular case of the Mooney-Rivlin material; for the incompressible case, $W$ is expressed as a function of an invariant of $\textbf{B}$:
\begin{equation}
    W=\frac{G}{2}(I_c-3),
\end{equation}
where $I_c = \rm{tr}(\textbf{F}^T\textbf{F})$, and $G$ is the transverse elasticity.
From above, the Cauchy stress tensor of the elastic material becomes as follows:
\begin{equation}
    \boldsymbol{\sigma} = G\textbf{B}.
\end{equation}
The details about the hyperelastic material can be found in e.g., \citet{bonet1997nonlinear} (chapter 5).

\section{Equivalent sand grain roughness}\label{asec: ks}
The procedure for determining the equivalent sand grain roughness $k_s$ is as follows. The semi-logarithmic mean velocity profile varies depending on the equivalent sand grain roughness $k_s$ as~\citep{nikuradse1933stromungsgesetze}:
\begin{equation}
    \bar{u}^+ = \frac{1}{\kappa}\ln\frac{y}{k_s}+8.5,
    \label{eq: Nk}
\end{equation}
where $\bar{u}^+$, $\kappa$, and $y$ are the streamwise mean velocity normalized by the friction velocity, the von Kármán constant, and the wall-normal distance, respectively. First, we introduce the wall-normal shift of the origin $d$ as done by \citet{Jackson_1981} for rough walls and \citet{Breugem2006The} for porous medium. 
Next, by replacing $y$ into $y+d$ \citet{okazaki2020turbulence}, we compare the previous Eq.~\ref{eq: Nk} with the following 
\begin{equation}
  \Bar{u}^{+} = \frac{1}{k + \Delta k} \log (\Tilde{y}+d)^+ + B - \Delta U^+,
 \label{eq: appendix fitting}
\end{equation}
so that we can obtain the length scale $k_s^+$ as
\begin{equation}
    \ln k_s^+ = \ln \Tilde{y} - \frac{\kappa}{k+\Delta k}\ln (\Tilde{y}+d) + \kappa(8.5-B+\Delta U^+).
    \label{eq: my ks}
\end{equation}
$k_s^+$ is a function of the wall-normal direction, and we take the values at $\Tilde{y}^+=d^+$ as representative $k_s^+$.

\section{The effect of the wall-normal fluctuations}\label{asec: porous}

\begin{figure}
    \centering
    \begin{minipage}[]{0.49\columnwidth}
        \centering
        \includegraphics[width=0.9\columnwidth]{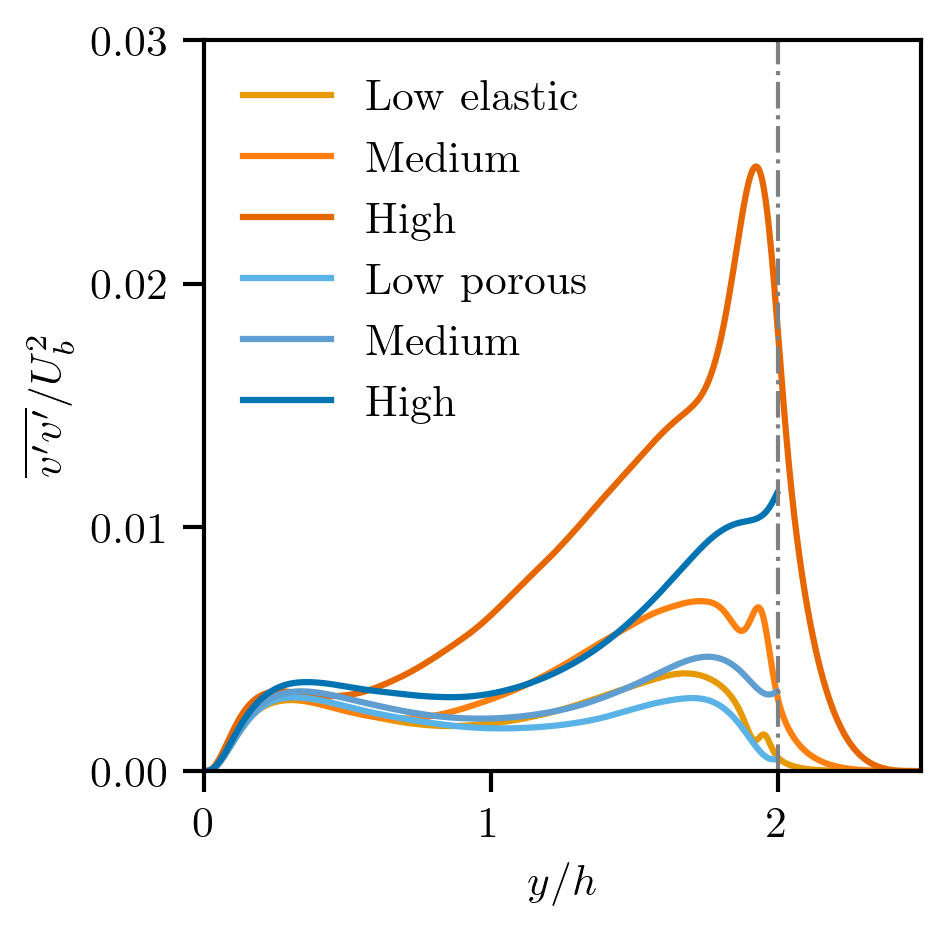}
    \end{minipage}
    \begin{minipage}[]{0.49\columnwidth}
        \centering
        \includegraphics[width=0.9\columnwidth]{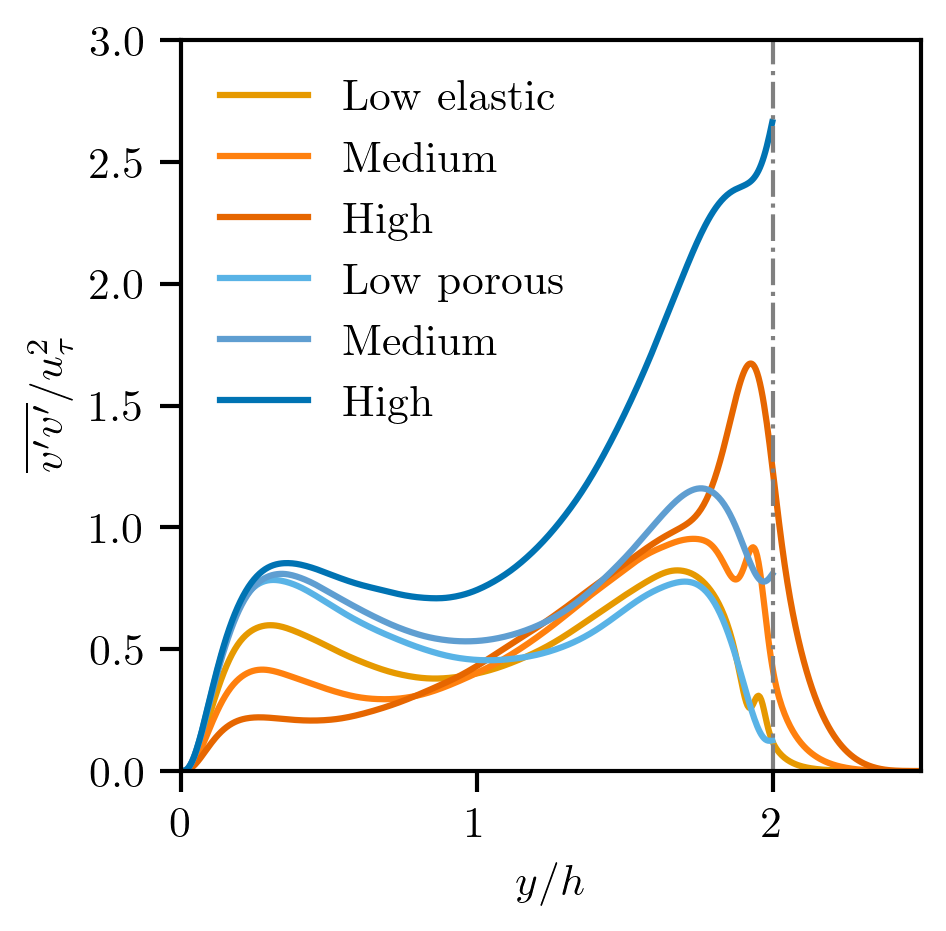}
    \end{minipage}
    \caption{The wall-normal components of the Reynolds stress tensor as a function of the wall-normal coordinate with the components normalized by (left) $U_b^2$ and (right) $u_\tau^2$, for the elastic (orange) and porous (blue) cases. The color brightness represents the level of wall-normal fluctuations at the wall, going from low to high (bright to dark). The gray dash-dotted line shows the position at $y/h = 2$, i.e., the interface between the fluid and solid phases.}
    \label{fig: porous validation}
\end{figure}

\begin{table}
  \begin{center}
  \def~{\hphantom{0}}
  \begin{tabular}{c|c|cc|cccccc}
        Wall &  Complexity &  $\beta(\rho U_b)$ &  $\beta(\rho u_\tau)$ &  $Re_\tau^w$ &  $Re_\tau$ &  $d/h$ &  $d^+$ &  $k+\Delta k$ &  $\Delta U^+$  \\[3pt]
       \hline
        Porous &  Low &  0.0221 &  0.0014 &  179.3 &  175.9 &  0.0~~ &  0.0~~ &  0.38~ &  0.1           \\
                  &  Medium &  0.0573 &  0.0036 &  178.2 &  178.7 &  0.0~~ &  0.0~~ &  0.38~ &  0.35  \\
                  &  High &  0.107~ &  0.0070 &  181.1 &  183.1 &  0.0~~ &  0.0~~ &  0.38~ &  0.6  \\
  \end{tabular}
  \caption{Summary of the parameters and flow characteristics in the porous cases. The table reports the permeability coefficient $\beta(\rho U_b)$ and $\beta(\rho u_\tau)$, the rigid wall friction Reynolds number $Re_\tau^w$, the porous wall friction Reynolds number $Re_\tau$, the wall-normal shift $d/h$ for the outer scale and $d^+$ for the inner scale, the modified von Kármán constant $k+\Delta k$, and the logarithmic shift $\Delta U^+$.}
  \label{tab: summary wall config and flow chara of e and p}
  \end{center}
\end{table}

\begin{figure}
  \centering
  \includegraphics[width=1\textwidth]{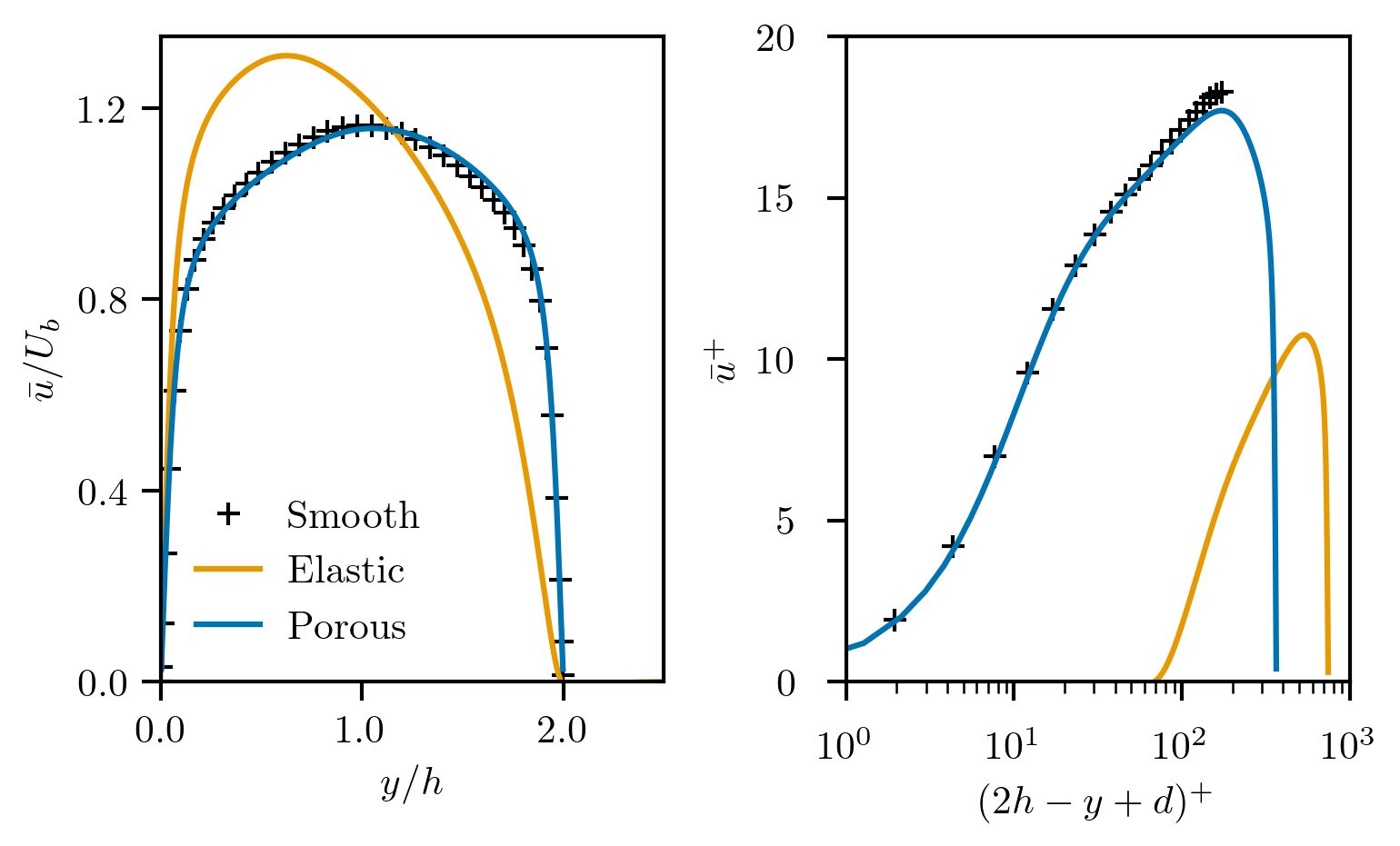}
  \caption{The streamwise velocity profile as a function of the wall-normal direction in the (left) outer and (right) inner scales, for the case with high wall-normal fluctuations at the wall. The line colors represent the wall types: orange (elastic) and blue (porous). The $+$ symbols correspond to the results of the classical turbulent channel flow by \citet{kim1987turbulence}.}
\label{fig: mean vel prof p}
\end{figure}

\begin{figure}
  \centering
  \includegraphics[width=1\textwidth]{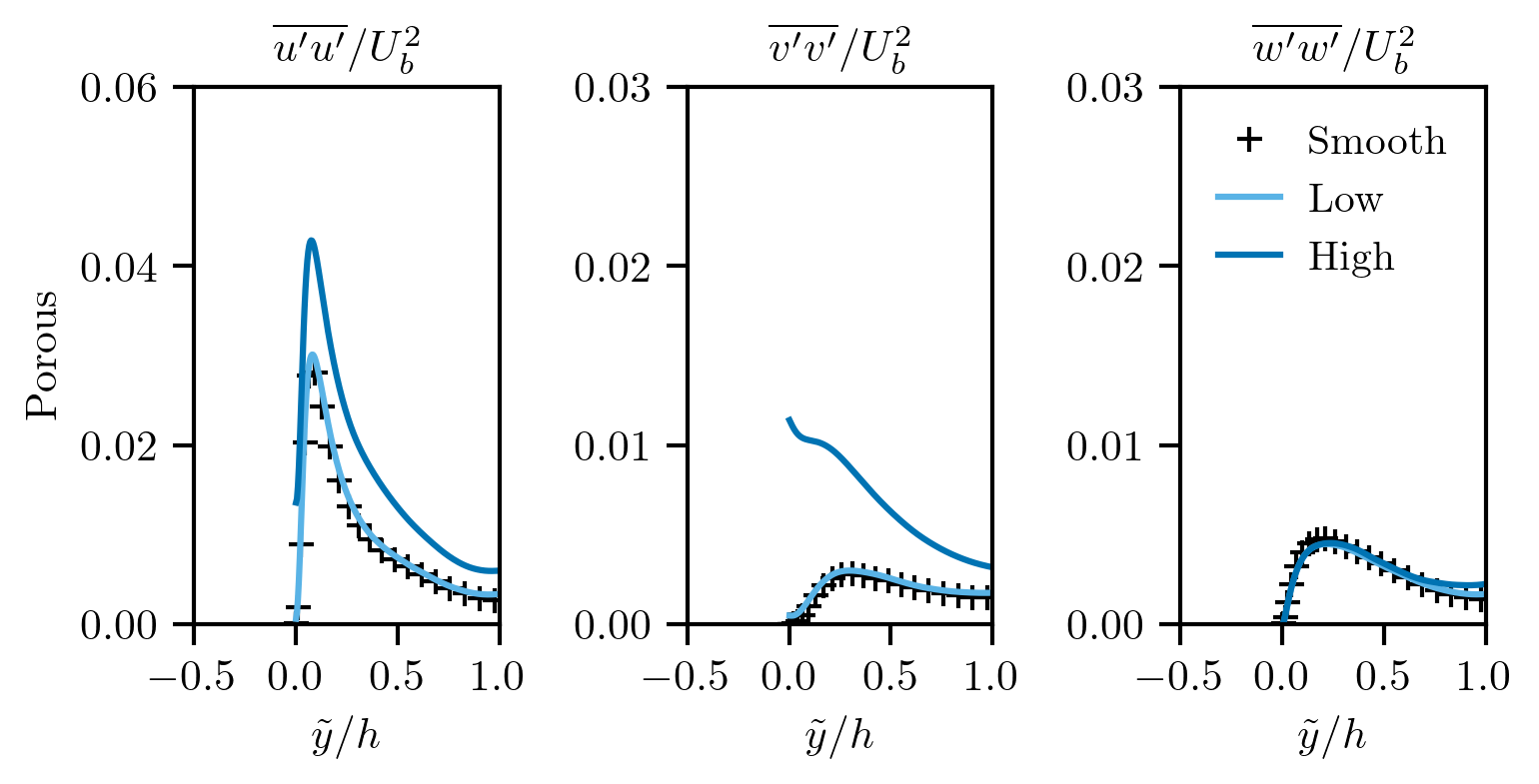}
  \caption{The diagonal terms of the Reynolds stress tensor normalized by $u_\tau$ for the porous wall as a function of the wall-normal distance. The color brightness indicates the level of $\beta$, going from low (bright) to high (dark). The $+$ symbols correspond to the results from~\citet{kim1987turbulence}.}
\label{fig: rss p}
\end{figure}

The dynamic movement of the elastic wall causes strong non-zero velocity fluctuations in the wall-normal direction, which can be linked to the reduction of coherency of the streamwise velocity fluctuations \citep{Rosti_Brandt_2017, Ardekani_Rosti_Brandt_2019}. Additionally, \citet{Breugem2006The,jimenez2001turbulent,kuwata2016lattice,rosti2018turbulent} reported a correlation between enhanced wall-normal fluctuations and the spanwise-coherent structures (rollers). Thus, in this section, we compare the results of the cases with elastic walls with those of model permeable walls (addressed as `porous' later), whose effects are implemented through imposed boundary conditions explained as follows. Instead of simulating the full porous medium, here we model the permeability effect by imposing the following boundary conditions at the interface between the fluid and the porous medium
\begin{equation}
  u = w = 0, \qquad v = -\beta p^\prime.
\end{equation}
In this effective boundary condition, the wall-normal velocity is assumed to be proportional to the local pressure fluctuations at the interface, and the amount of permeability can be controlled by varying the proportionality coefficient $\beta$. The classical impermeability condition, $v = 0$, is recovered when $\beta=0$, while $\beta \rightarrow \infty$ represents an unrestrained wall-normal velocity and zero pressure fluctuations on the wall. These boundary conditions have been proposed and used before to study turbulent channel flows over porous surfaces by \citet{jimenez2001turbulent}, who also reported spanwise organized structures. In our study, we consider three cases, summarized in Table~\ref{tab: summary wall config and flow chara of e and p}, with increasing levels of permeability controlled by the parameter $\beta$, whose values are chosen such that the wall-normal fluctuations in bulk units, on average, are similar at the interface of the porous and the corresponding elastic cases.
We confirm this in Figure \ref{fig: porous validation}, where we show the profile of the wall-normal components of the Reynolds stress tensor for all elastic (orange) and porous (blue) cases. In Figure~\ref{fig: porous validation} (left), the velocity fluctuations are scaled by $U_b^2$, while in Figure~\ref{fig: porous validation} (right), they are scaled by $u_\tau^2$. From Figure~\ref{fig: porous validation} (left), in the low and medium cases, the velocity values of elastic and porous walls at the interface are of comparable intensity. Note that the average fluctuations for porous cases were selected to be the same as elastic cases, while the rest of the profile is not forced and arises naturally from the evolution of the governing equations. Furthermore, we report in Figure~\ref{fig: porous validation} (right) the same data in plus units, which shows a quantitative mismatch of the velocity fluctuation at the wall interface. However, the qualitative trend with the complexity parameter ($G$ and $\beta$) remains similar, i.e., the fluctuations increase in magnitude with increasing the complexity degree for both the elastic and porous cases.

Table~\ref{tab: summary wall config and flow chara of e and p} reports the flow characteristics for the porous cases. Clearly, for the present set of parameters, there is no significant variation in the friction Reynolds number $Re_\tau$ of the porous wall, which remains similar to those of the canonical turbulent channel flow, except for a quite mild increase. Figure~\ref{fig: mean vel prof p} exhibits the mean velocity profiles in both outer and inner scales for the high wall elasticity/permeability case (note that the results for the elastic walls are the same as discussed in the previous sections). As expected from the values of $Re_\tau$, the distribution of the mean velocity profile is only slightly affected by the present permeable walls. Similarly, with increasing wall permeability, the Reynolds stresses in the streamwise and wall-normal directions are enhanced close to the permeable wall, while the ones in the spanwise direction remain almost unaltered, as shown in Figure~\ref{fig: rss p}.

\begin{figure}
  \centering
  \includegraphics[width=1\textwidth]{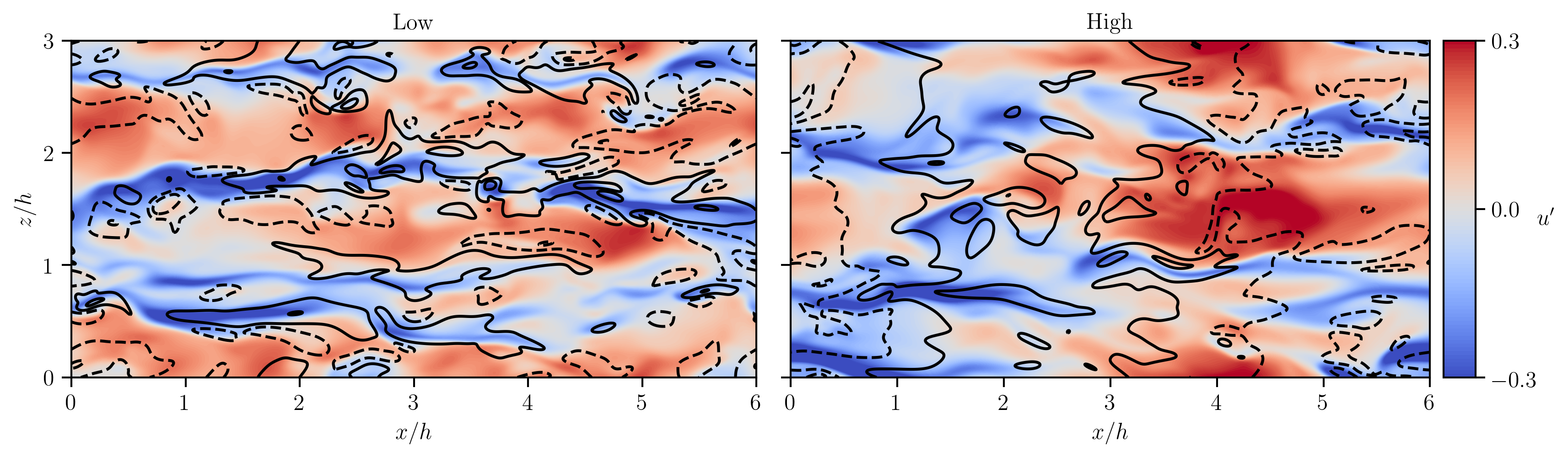}
  \caption{Contour fields of the streamwise velocity fluctuations $u^\prime$ (color map) and wall-normal fluctuations $\Tilde{v}^\prime$ (isolines) at $2h-y=0.2h$ for the porous case. The color map for $u^\prime$ ranges from $-0.3$ (blue) to $0.3$ (red), while the contour of $\Tilde{v}^\prime$ goes from $-0.15$ to $0.15$, with increments of $0.01$. Dashed lines are used for negative contours. The left and right columns correspond to the low and high permeability cases, and the flow direction is from left to right.}
\label{fig: streak p}
\end{figure}

\begin{figure}
  \centering
  \includegraphics[width=0.8\textwidth]{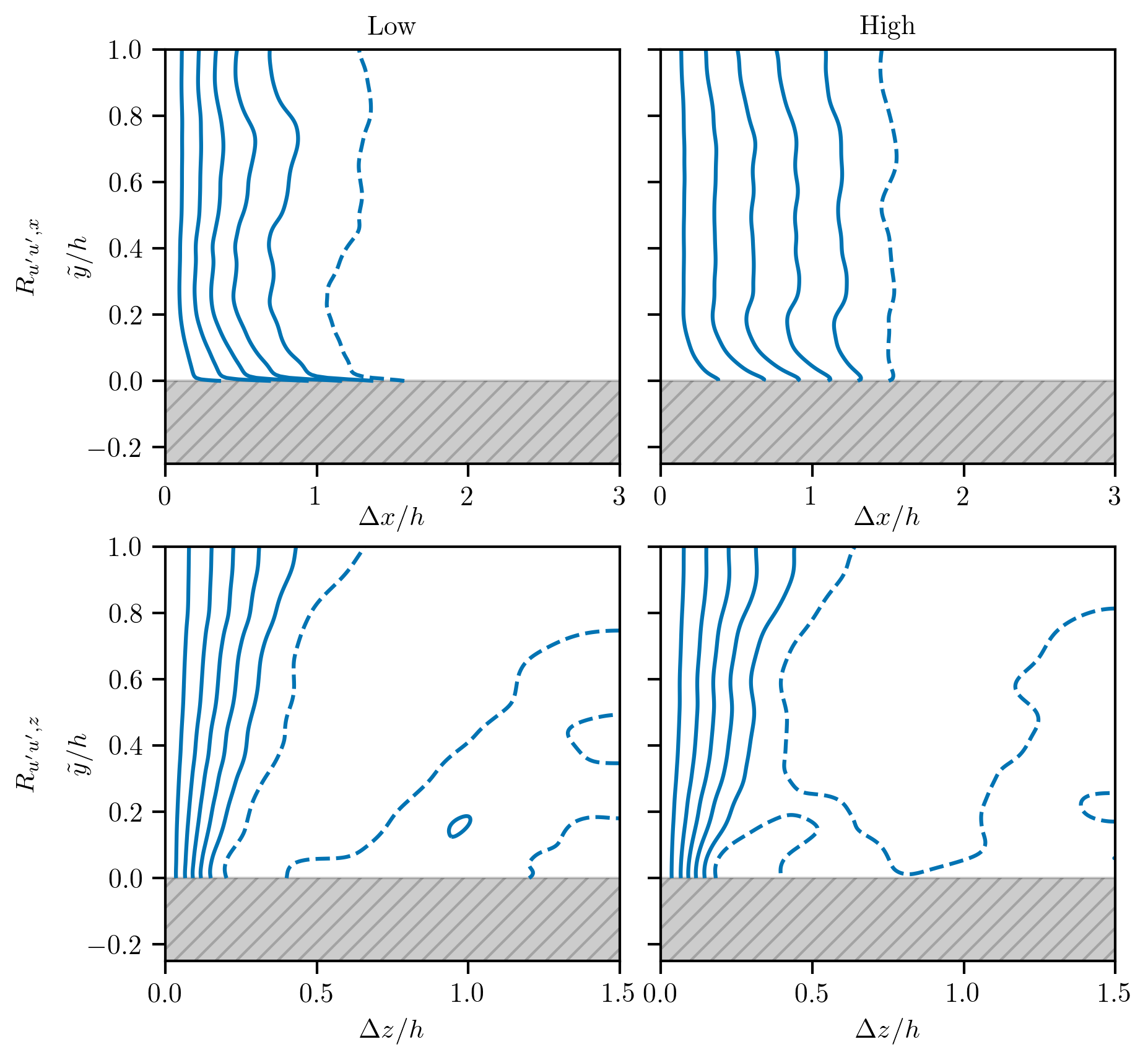}
  \caption{One-dimensional autocorrelation of the streamwise velocity fluctuations in the (top) streamwise and (bottom) spanwise directions. The background colors and line representation are the same as Figure~\ref{fig: space Ruux}.}
  \label{fig: space Ruuxp}
\end{figure}

To investigate the turbulent coherent structures, we look again at the instantaneous streamwise velocity fluctuations in the wall-parallel plane at $2h-y=0.2h$, shown in Figure~\ref{fig: streak p}. We can observe that the low and high-speed streaks are retained, with only a minor increase in the spanwise length for the high wall permeability case. This tendency is also observed in the one-dimensional autocorrelation functions of the streamwise velocity fluctuations, plotted as a function of the streamwise and spanwise spacings in Figure~\ref{fig: space Ruuxp}; at $2h-y=0.2h$, the correlation length within both directions increases for increasing permeability. Additionally, for high wall-normal fluctuations, the distinctive wavy distribution previously observed for the elastic cases in Figure~\ref{fig: space Ruux} is not present for the porous case.

We briefly address here some of the differences and similarities between the current observations of the porous configurations compared to \citet{jimenez2001turbulent}, noting that the permeability of the interface is not the same in the two works. \citet{jimenez2001turbulent} report large spanwise coherent structures, with a strong spanwise correlation of the streamwise velocity fluctuations. While our highly porous case shows a stronger increase in correlation along the streamwise direction compared to the spanwise one, still the coherency in the spanwise direction increases as we increase the wall permeability, as shown in Figure~\ref{fig: streak p} and in agreement with their results. Similarly, \citet{jimenez2004turbulent} report low-velocity streaks in the regions of blowing and high-velocity streaks in the regions of suction, an effect of the vertical disturbances of the permeable wall; in Figure~\ref{fig: streak p}, we have depicted the wall-normal velocity component superimposed with the streamwise velocity streaks for our results, and similar conclusions can be drawn.

Overall, the porous cases show that, in spite of the similar wall-normal velocity fluctuations at the boundary, the flow behaviour between porous and elastic walls stays different, suggesting these do not solely dictate the near-wall (elastic) dynamics. Note, however, that the porous media have been simulated only with the simplest model available in the literature, where all the effects are concentrated into a boundary condition depending on a single parameter.





\backsection[Acknowledgements]{This research was supported by the Okinawa Institute of Science and Technology Graduate University (OIST) funding to MER from the Cabinet Office, Government of Japan. The authors acknowledge the computer time provided by the Scientific Computing \& Data Analysis section of the Core Facilities at OIST.}


\backsection[Declaration of interests]{The authors report no conflict of interest.}


\backsection[Author ORCIDs]
{\\
M. Koseki, https://orcid.org/0000-0002-3550-4378 \\
M. S. Aswathy, https://orcid.org/0000-0003-4586-7364 \\
M. E. Rosti, https://orcid.org/0000-0002-9004-2292}


\bibliographystyle{jfm}

\end{document}